\documentclass[pra,twocolumn,aps,superscriptaddress]{revtex4-1}
\usepackage{amsmath}
\usepackage{amssymb}
\usepackage{graphicx}
\usepackage[usenames,dvipsnames]{color}
\usepackage{braket}
\usepackage{bm}
\usepackage{times}
\usepackage{hyperref}
\usepackage{pdfpages}
\usepackage{float}
\usepackage{lipsum}
\usepackage{soul}
\usepackage[normalem]{ulem}
\usepackage[dvipsnames]{xcolor}
\hypersetup{
  colorlinks=true,
  citecolor=blue,
  linkcolor=blue,
  urlcolor=blue}
\usepackage{tikz}
\usetikzlibrary{decorations.markings}
\usetikzlibrary{decorations.pathmorphing,snakes}
\usetikzlibrary{arrows.meta}
\tikzset{middlearrow/.style={
        decoration={markings,
            mark= at position 0.5 with {\arrow{#1}} ,
        },
        postaction={decorate}
    }
}

\makeatletter
 \AtBeginDocument{\let\LS@rot\@undefined}
\makeatother

\begin{document}
\title{Miscibility-immiscibility transition of strongly interacting bosonic 
mixtures in optical lattices}

\author{Rukmani Bai}
\email{rukmani.bai@itp3.uni-stuttgart.de}
\affiliation{Institute for Theoretical Physics III and Center for 
             Integrated Quantum Science and Technology, University of 
	     Stuttgart, 70550 Stuttgart, Germany}
\affiliation{Institut f\"ur Theoretische Physik, Leibniz Universit\"at Hannover,
Appelstrasse 2, D-30167 Hannover, Germany} 	     
\author{Soumik Bandyopadhyay}
\email{soumik.bandyopadhyay@unitn.it}
\affiliation{Pitaevskii BEC Center, CNR-INO and Dipartimento di Fisica, Universit\`a di Trento, Via Sommarive 14, Trento, I-38123, Italy}
\affiliation{INFN-TIFPA, Trento Institute for Fundamental Physics and Applications, Trento, Italy}

\begin{abstract}
Interaction plays key role in the mixing properties of a multi-component system. The miscibility-immiscibility transition (MIT) in a weakly interacting mixture of Bose gases is predominantly determined by the strengths of the intra and inter-component two-body contact interactions.
On the other hand, in the strongly interacting regime interaction induced processes become relevant. Despite previous studies on bosonic mixtures in optical lattices, the effects of the interaction induced processes on the MIT remains unexplored.
In this work, we investigate the MIT in the strongly interacting phases of two-component bosonic mixture trapped in a homogeneous two-dimensional square optical lattice. Particularly we examine the MIT condition when both the components are in superfluid (SF), one-body staggered superfluid (OSSF), or supersolid (SS) phases. 
Our study uncovers that MIT  condition is significantly shaped by the interplay of competing non-local intra- and inter-component density-induced tunneling effects, as well as off-site interactions. Notably, we demonstrate that the MIT condition for the staggered superfluid phase exhibits an inequality that is inverted compared to the conventional MIT condition associated with superfluid or supersolid phases driven by local contact interactions. In addition, we present the phase diagram of the Bose-Hubbard Model incorporating non-local processes, derived using a site-decoupling mean-field approach with the Gutzwiller ansatz.
Our study contributes to the better understanding of miscibility properties of multi-component systems in the strongly interacting regime. 

\end{abstract}



\maketitle


\section{Introduction}
\label{sec1}
Miscibility-Immiscibility transition (MIT) is ubiquitous in nature. Depending
on the interaction of microscopic constituents, like van der Waals force between
the molecules, some fluids may mix with each other whether others do not.
Physical properties of these multi-component fluids, such as, temperature,
density, pressure, etc., and external forces are conventionally known to influence 
the mixing properties. Intrigued by its simplicity with vast range of interdisciplinary 
implications \cite{jacopo_11}, MITs have been explored extensively in quantum fluids 
particularly with trapped ultracold gases of atoms in the weakly interacting regime 
\cite{hall_98, mertes_07, papp_08, tojo_10, mccarron_11, wacker_15, wang_16, eto_16}.
These multi-component systems can be composed of atoms of different elements 
\cite{modugno_01, modugno_02, lercher_11, mccarron_11, pasquiou_13, wacker_15, wang_16, 
luca_22}, different
hyperfine states \cite{myatt_97, hall_98, stamper_98, stenger_98, 
maddaloni_00, delannoy_01, sadler_06, mertes_07, anderson_09, tojo_10, eto_16, fava_18, 
farolfi_20, farolfi_21, cominotti_22} or 
isotopes \cite{papp_08, handel_11, sugawa_11a} of same element, or even elements 
obeying different quantum statistics \cite{barbut_14, desalvo_17, desalvo_19}.
In a multitude of previous studies, effects of
system characteristics, e.g., confining potentials \cite{gautam_10_2, gautam_11, richaud_19, richaud_19_1}, 
number-imbalance \cite{pattinson_13, lee_16, wen_20}, 
temperature \cite{kwangsik_07, roy_14_1, roy_15_2, lee_16, suthar_17, ota_19, 
arko_21, spada_23}, 
impurity and topological defects \cite{bandyopadhyay_17, kuopanportti_19} on the MITs 
of superfluids, also referred to as phase-separation 
\cite{ao_98, timmermans_98, pu_98}, have been 
explored. In such weakly interacting systems, the MITs are mainly determined by the 
competition between the short-range interactions of the components 
\cite{ho_96, ao_98, pu_98, pethick_2002}.

By restricting the atoms to interact in the wells of
deep optical lattices, the effective strength of the interaction can be
made significantly strong admitting novel quantum phase transitions 
\cite{oliver_06, bloch_08, lewenstein_12, gross_17}.
In this regime, interaction induced processes are known to influence the
physical characteristics of the system, and in some cases can stabilize new
quantum phases resulting significant modifications to the phase diagrams
\cite{mazzarella_06, dutta_15}. 
Here, we explore the role of different interactions on the MITs
of bosonic multi-component systems in the strongly interacting regime.

For past two decades, the trapped ultracold atoms have been in the forefront
of quantum simulation and exploration of synthetic quantum matters 
\cite{lewenstein_12, bloch_12, gross_17}. Excellent experimental control, 
e.g., near perfect state initialization \cite{yang_20_1, yang_20_2, ebadi_21} 
to the readout at the level of 
individual atoms \cite{bakr_09, kuhr_16}, has made these systems popular 
platforms to probe plethora of models ranging from condensed-matter physics 
\cite{jaksch_05, lewenstein_07, zhang_18, hofstetter_18, tarruell_18}, 
high-energy physics \cite{wiese_13, aidelsburger_16, zohar_16, halimeh_23}, 
quantum information \cite{monroe_02, jaksch_04, garcia_05}, 
nuclear physics \cite{gezerlis_08, giancarlo_18}, 
cosmology \cite{uwe_04, hung_13, schmiedmayer_13, chatrchyan_21} 
to quantum gravity \cite{wei_21, uhrich_23}. Under the assumption of lowest-band
approximation, the minimal model which can capture the physics of an optical
lattice with ultracold bosonic atoms is the bosonic version of the Hubbard model,
a.k.a. the Bose-Hubbard model (BHM) \cite{hubbard_63, fisher_89, jaksch_98}, where 
the tunneling between the nearest-neighbour wells and only the onsite interaction
is considered \cite{fisher_89, jaksch_98}. This model has been tremendously 
successful in describing the Superfluid (SF) to Mott-insulator (MI)
transition \cite{greiner_02, bakr_10}. However, the model demands an extension
through the inclusion of other processes, such as offsite interactions,
density-induced and pair tunneling, etc. \cite{mazzarella_06}, when the interaction 
between the atoms is stronger yielding multi-band processes and 
long-range interactions relevant~\cite{sowinski_12, Luhmann_12, Jurgensen_12, maik_13, Jurgensen_14, Jurgensen_15, 
Luhmann_16, dutta_15}. 
Such extended versions of the BHM are known to exhibit other phases, for example, density-wave (DW)~\cite{mazzarella_06}, supersolid~\cite{sengupta_05,scarola_05,ng_08,iskin_11,suthar_20_1}, staggered superfluids and supersolids~\cite{John_19,suthar_20_2,suthar_22, baier_16}, etc. 

In general, the long-range interactions and density induced tunneling (DIT)
processes have implications in fast-information scrambling and 
quantum-many body chaos \cite{chowdhury_22, bandyopadhyay_23}, spin glasses 
and quantum annealing algorithms \cite{panchenko_15, lechner_15} to the 
understanding properties of superconductors and 
magnetic materials \cite{pu_01, paz_13}. 
In modeling optical lattice systems, non-local interactions and DIT are often limited to nearest-neighbor (NN) sites to capture leading-order effects. For atoms or molecules with large dipole moments, these processes become relevant~\cite{goral_02, danshita_09, capogrosso_10, sowinski_12, maik_13, bandyopadhyay_19}. Motivated by their intriguing features and recent experimental advances with bosonic dipolar atoms~\cite{baier_16} and molecules~\cite{lukas_17}, we investigate how NN interactions and DIT influence the MITs in two-component bosonic mixtures in optical lattices.
Specifically, we demonstrate that both DIT and NN interactions significantly influence the MIT in the superfluid and supersolid phases. Most notably, we uncover that the MIT condition becomes reversed in the staggered superfluid phase—revealing a striking departure from conventional behavior and underscoring the profound impact of these interaction induced processes.

The key result of this manuscript is summarized in the Fig.~\ref{fig_mit} 
(see Sec.\ref{sec_mitstrong}), which we subsequently extend to large range of parameter domains. For better readability, we organize the manuscript as follows: In Sec. \ref{sec_model}, 
we discuss the model Hamiltonian, mean-field methodology and quantum phases hosted by the model. 
In Sec. \ref{sec_mitstrong} we 
discuss the MIT of SF, OSSF and SS phases with respect to the DIT processes and NN interactions.
We extend our results by examining the influence of these processes on the immiscibility 
condition in Sec. \ref{sec_brod_param}. In this section, we additionally present the 
phase diagrams of the model with DIT processes (Sec. \ref{sec_phdiag_bhm_dit}) 
and both DIT processes and NN interactions (Sec. \ref{sec_phdiag_bhm_dit_nn}).
Finally, we summarize our key findings and highlight experimental feasibility 
of our observations in Sec. \ref{sec_summary}.
%
\section{Model}
\label{sec_model}
\begin{figure}[t]
\includegraphics[width=7.5cm]{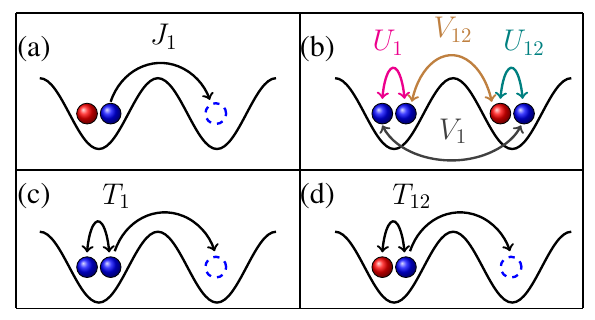}
\caption{Schematic illustration of different terms in the model 
         Hamiltonian [see Eqs. (\ref{tbhm}) and (\ref{hamil_terms})] of  
         two-component bosonic mixture in optical lattice potential. 
         The red and blue color balls represent the particles of 
         different components. The $J$s, $V$s, 
         $T$s and $U$s denote the strengths of the nearest-neighbour 
         hoppings, interactions, density induced tunnelings and onsite 
         interactions, respectively, with the subscripts marking 
         intra and inter-component processes. }
\label{fig_schematic}
\end{figure}

To study the MIT, we consider a system comprising two-component mixture of bosonic  
particles trapped in the potential of a two-dimensional (2D) optical lattice comprising $L\times L$ sites. 
Under the second quantization framework, the grand canonical Hamiltonian of such a system is given by
\begin{equation}
\label{tbhm}  
\hat{H} = \hat{H}^{{\rm TBH}} + \hat{H}^{{\rm DIT}} + \hat{H}^{{\rm NN}},
\end{equation}
where 
\begin{eqnarray}
\label{hamil_terms}
&\hat{H}^{{\rm TBH}}& 
= -\sum_{\langle i, j\rangle, \kappa}\left( J_\kappa\hat{b}_{i,\kappa}^\dagger
    \hat{b}_{j,\kappa} + {\mathrm{H.c.}}\right)- \mu_\kappa\sum_{i} \hat{n}_{i,\kappa}  \nonumber\\
	      &&+\sum_{i,\kappa} \frac{U_{\kappa}}{2} \hat{n}_{i,\kappa} (\hat{n}_{i,\kappa}-1)  
	        + U_{12}\sum_{i} \hat{n}_{i,1}\hat{n}_{i,2},\nonumber\\
&\hat{H}^{{\rm DIT}}& = \sum_{\langle i, j\rangle , \kappa} 
              \left[T_{\kappa} 
	      \hat{b}_{i,\kappa}^{\dagger} 
	      \left(\hat{n}_{i,\kappa} + \hat{n}_{j,\kappa}\right)\hat{b}_{j,\kappa}\right.\nonumber\\
	      && + T_{\kappa,3-\kappa} \left.\hat{b}_{i,\kappa}^{\dagger}
	      \left(\hat{n}_{i,3-\kappa} + \hat{n}_{j,3-\kappa}\right)\hat{b}_{j,\kappa} 
         + {\mathrm{H.c.}}     \right],\nonumber\\
&\hat{H}^{{\rm NN}}& = \sum_{\langle i, j\rangle , \kappa}  
	            V_{\kappa} \hat{n}_{i,\kappa} \hat{n}_{j,\kappa} 
        + V_{\kappa, 3 - \kappa} \hat{n}_{i,\kappa} \hat{n}_{j,3-\kappa}. 
\end{eqnarray}
In  Eqs. (\ref{tbhm}) and (\ref{hamil_terms}), the first term, $\hat{H}^{\rm TBH}$,
corresponds to the standard two-component BHM Hamiltonian, 
where $\kappa = 1$, $2$ stands as the component index, and $i$, $j$ denote
the lattice site indices of a square lattice with
the notation $\langle \bullet \rangle$ representing NN sites. 
The local creation, annihilation and number operators of the $i$th site 
are denoted by $\hat{b}^{\dagger}_{i,\kappa}$, $\hat{b}_{i,\kappa}$ and 
$\hat{n}_{i,\kappa}$, respectively. $J_\kappa$ corresponds to the strength of the 
hopping process of particles within NN sites, $U_{\kappa}$ is 
the strength of the onsite repulsive contact interaction of individual 
component, and $U_{12}$ is the inter-component onsite interaction strength. We consider all these strengths positive, uniform and isotropic, and 
probe the regime when $J_1 = J_2 = J$ is sufficiently smaller than 
$U_1 = U_2 = U$.

The second term, $\hat{H}^{\rm DIT}$, corresponds to DIT processes of the two-components.
Here, $T_{\kappa}$ and $T_{12} = T_{21}$ denote  the intra and inter-component 
DIT strengths, respectively.
The third term, $\hat{H}^{\rm NN}$, represents the density-density interaction between the particles in the NN sites. In the equation, $V_{\kappa}$ and $V_{12} = V_{21}$ are the 
strengths of intra and inter-component NN interactions.
In Eq.~(\ref{hamil_terms}),
the chemical potential $\mu_{\kappa}$ enforces the particle number conservation 
in the grand-canonical ensemble, and determine the fillings of individual 
component in the lattice system. 
In the present work, we fix $\mu_1 = \mu_2 = \mu$. 

In Fig.~\ref{fig_schematic}, we schematically illustrate the different 
processes described by the model Hamiltonian in Eqs. (\ref{tbhm}) and 
(\ref{hamil_terms}). 
 
Furthermore, a detailed description of these processes, along with a discussion of the resulting phases supported by the model, is provided in Appendix~\ref{append1}.
The Hamiltonian can additionally be extended via 
incorporating pair hopping processes yielding novel phases of particle pairs,
e.g., pair superfluids and supersolids. But, such processes are weaker than 
the DIT processes for systems with interacting particles
\cite{mazzarella_06, John_19}. Even for the systems comprising dipolar particles the 
pair hopping process is weaker than the DIT and NN interactions 
\cite{sowinski_12}. Therefore, in this work we restrict the extension of the 
BHM considering the latter two processes.

\subsection{Single-site mean-field Hamiltonian and energy}
\label{ss_mfhamil_energy}
In this work, we employ single-site mean-field theory to obtain the ground states of the Hamiltonian in Eqs. (\ref{tbhm}) and (\ref{hamil_terms}) \cite{rokhsar_91, sheshadri_93, anufriiev_16, bai_18, bandyopadhyay_19, bai_20}. The aim of this framework is to express the system Hamiltonian as a sum of the local Hamiltonians of individual sites. For this, the non-local terms in Eq. (\ref{hamil_terms}) are decoupled through the introduction of the mean fields of different operators. To do so, the annihilation, creation, and number operators of each site are decomposed into corresponding mean fields and fluctuation operators. That is, $\hat{A}_{i,\kappa} = \langle\hat{A}_{i,\kappa}\rangle + \delta\hat{A}_{i,\kappa}$, where 
$\langle\hat{A}_{i,\kappa}\rangle$ and $\delta\hat{A}_{i,\kappa}$
are the mean field and fluctuation operator of a representative single-site operator $\hat{A}_{i,\kappa}$. Then, to leading order in the fluctuations, different terms of Eq. (\ref{hamil_terms}) are obtained (see Eq. (\ref{mean-field-terms})).
Thus, the mean-field version of the Hamiltonian can be written as a sum of local Hamiltonians of the sites:
\begin{equation}
	\hat{H}^{\rm MF} = \sum_{i}\hat{h}_{i}^{\rm MF} = \sum_{i}\hat{h}_{i}^{{\rm TBH}} 
	+ \hat{h}_{i}^{{\rm DIT}} + \hat{h}_{i}^{{\rm NN}}.
\label{hamil_ss_tbec}
\end{equation}
The explicit form of this Hamiltonian for a 2D square lattice is provided in Appendix~\ref{append1} (see Eq. (\ref{ham_ss_tbec_a})).

The single-site Hamiltonian matrix can be constructed with respect to the local Fock basis $\{|m_1,m_2\rangle\}$, where $m_\kappa$ is the occupation number of the $\kappa$-th component, and then the matrix can be diagonalized to obtain the ground state of the single-site Hamiltonian. From this, the ground state of the mean-field Hamiltonian  in Eq. (\ref{hamil_ss_tbec}) can be constructed by employing the Gutzwiller ansatz \cite{jakub_05,yan_17,niederle_13,jaksch_02,damski_03_1,yan_17_1,yan_17_2,kaur_24}, where the ground state of the system is expressed as the tensor product of ground states of all the local Hamiltonians, i.e., 
\begin{equation}
 |\Psi_{\rm GW}\rangle = \prod_{\otimes i} 
  |\psi\rangle_{i} = \prod_{\otimes i}\sum_{m_1, m_2 }f^{i}_{m_1, m_2} 
                       |m_1, m_2\rangle_{i}.
  \label{gw_2s}
\end{equation}
Here, $|\psi\rangle_i$ is the ground state of the mean-field Hamiltonian of the $i$-th site, and 
the coefficients, $f_{m_1, m_2}^{i}$, 
satisfy the normalization condition 
$\sum_{m_1,m_2} |f^{i}_{m_1,m_2}|^2$ = 1. 

The local SF order parameter and density of the first component can be calculated from the Gutzwiller ground state as 
  \begin{eqnarray}
    \phi_{i,1}&=& \langle\Psi_{\rm GW}|\hat{b}_{i,1}|\Psi_{\rm GW}\rangle 
            = \sum_{m_1,m_2}\sqrt{m_1} 
	    f^{i*}_{m_1-1, m_2} f^{i}_{m_1,m_2},\nonumber\\
   \rho_{i,1} &=& \langle\Psi_{\rm GW}|\hat{n}_{i,1}|\Psi_{\rm GW}\rangle
            = \sum_{m_1, m_2} m_1 |f^{i}_{m_1,m_2}|^2.
    \label{gw_phi}           
  \end{eqnarray}
  These quantities can be calculated for the second component in the similar manner. The order parameters related to the density induced transport can be calculated as 
  \begin{eqnarray}
    \eta_{i,1}&=& \langle\Psi_{\rm GW}|\hat{n}_{i,1}\hat{b}_{i,1}|\Psi_{\rm GW}\rangle\nonumber\\
	&=& \sum_{m_1, m_2}\sqrt{m_1}(m_1 -1) 
	      f^{i*}_{m_1-1, m_2}f^{i}_{m_1,m_2},
              \label{gw_phi2s_1}\nonumber\\
    \eta_{i,12}&=& \langle\Psi_{GW}|\hat{n}_{i,1}
	      \hat{b}_{i,2}|\Psi_{\rm GW}\rangle\nonumber\\ 
	    &=&\sum_{m_1, m_2}m_1\sqrt{m_2} 
	      f^{i*}_{m_1, m_2-1}f^{i}_{m_1,m_2},
    \label{gw_eta}    
  \end{eqnarray}
and similarly $\eta_{i,2}$ and $\eta_{i,21}$ can be evaluated. 

A few important remarks about these order parameters are crucial for interpreting our results. The Hamiltonian in Eqs. \eqref{tbhm} and \eqref{hamil_terms} is time reversal invariant. Due to the presence of only real parameters, the Hamiltonian matrix is real-symmetric, yielding real eigenstates. The corresponding single-site Hamiltonian is real-symmetric as well. This implies $f^{i}_{m_1, m_2}\in\mathbb{R}$ in Eq. \eqref{gw_2s}. Then, quantities in Eqs. \eqref{gw_phi}-\eqref{gw_eta} are real. 
Additionally, invoking the single-site mean-field decomposition, the order parameters for density induced transport can be written as  
\begin{equation}
\eta_{i,\kappa} \approx \langle\hat{n}_{i,\kappa}\rangle\hat{b}_{i,\kappa} + \hat{n}_{i,\kappa} \langle\hat{b}_{i,\kappa}\rangle -\langle\hat{n}_{i,\kappa}\rangle \langle\hat{b}_{i,\kappa}\rangle.
\label{mf_eta}
\end{equation}
Even though this decomposition is formally correct, its evaluation with respect to the state in Eq. \eqref{gw_2s} is subtle. In our analysis, being onsite term, we do not perform this decomposition and account it exactly. However, 
Eq. \eqref{mf_eta} demonstrates that to the leading order of the fluctuations, the expectation value of these order parameters 
$\langle\eta_{i,\kappa}\rangle\approx\langle\hat{n}_{i,\kappa}\rangle \langle\hat{b}_{i,\kappa}\rangle$. Now, due to  $\rho_{i,\kappa}\geq 0$, $\textrm{sgn}(\eta_{i,\kappa}) = \textrm{sgn}(\phi_{i,\kappa})$. Similarly, $\textrm{sgn}(\eta_{i,\kappa,3-\kappa}) = \textrm{sgn}(\phi_{i, 3-\kappa})$. We also consistently observe this characteristic in our numerical analysis.

Furthermore, following Eq. (\ref{ham_ss_tbec_a}), we can cast the single-site mean-field energy function in terms of these mean-fields as $\mathcal{E}_{i} = \langle\hat{h}_{i}^{\rm MF}\rangle = \mathcal{E}^{\rm HOP}_{i} + \mathcal{E}^{\rm INT}_{i} + \mathcal{E}^{\rm DIT}_{i} + \mathcal{E}^{\rm NN}_{i}$. Here the energy
due to hopping of the particles is
\begin{eqnarray} 
\mathcal{E}^{\rm HOP}_{i} =  \sum_{\kappa} - J_\kappa\phi_{i,\kappa}\bar{\phi}^{*}_{i, \kappa},
\end{eqnarray} 
the onsite interaction energy is given by 
\begin{eqnarray} 
\mathcal{E}^{\rm INT}_{i} =  
  \sum_{\kappa}\frac{U_{\kappa}}{2}\langle\hat{n}_{i,\kappa}
              \left (\hat{n}_{i,\kappa}-1\right)
              \rangle
    + U_{12}\langle\hat{n}_{i, 1} \hat{n}_{i, 2}\rangle,\nonumber\\
\end{eqnarray}
the energy arising from the DIT processes is 
\begin{eqnarray} 
\mathcal{E}^{\rm DIT}_{i} &=&
\sum_{\kappa}\bigg\{T_{\kappa}
        \Big[\eta_{i,\kappa}\bar{\phi}_{i,\kappa}^{*}
        +\phi_{i,\kappa}\bar{\eta}_{i,\kappa}^{*}\Big] 
        \nonumber\\  
	      &+& T_{\kappa, 3-\kappa} 
        \Big[\eta_{i,3-\kappa,\kappa}
        \bar{\phi}_{i, \kappa}^{*} 
        + \phi_{i,\kappa}\bar{\eta}_{i, 3-\kappa, \kappa}^{*} \Big] \bigg\},\nonumber\\
\end{eqnarray}
and the NN interaction energy is 
\begin{eqnarray} 
 \mathcal{E}^{\rm NN}_{i} =
\sum_{\kappa} \bigg[ \frac{V_{\kappa}}{2} \rho_{i,\kappa}\bar{\rho}_{i, \kappa} 
+  \frac{V_{\kappa, 3-\kappa}}{2} \rho_{i,\kappa} \bar{\rho}_{i, 3-\kappa}\bigg].
\end{eqnarray} 

For discussions that will soon become relevant, it is important to note that in quantum phases exhibiting a checkerboard pattern in the mean-fields, the single-site energy can be conveniently expressed using a bipartite lattice description. This spatial modulation can be interpreted as the system comprising two interlaced sublattices (SL), labeled $A$ and $B$, arranged with a one-site periodicity along both spatial directions. In this configuration, every nearest-neighbor (NN) site of a lattice site in sublattice-$A$ belongs to sublattice-$B$, and vice versa. Accordingly, the spatial distribution of the order parameter $\phi_{i,\kappa}$ can be compactly represented as $\Phi_{\kappa}^{\rm SL} = (\phi_{\kappa}^{A}, \phi_{\kappa}^{B})$, where $\phi_{\kappa}^{A}$ and $\phi_{\kappa}^{B}$ denote the values of the superfluid (SF) order parameter on sublattices A and B, respectively. A similar notation can be adopted for other order parameters as well.
Additionally, as previously explained, we omit the complex conjugation symbols (i.e., the asterisk notation) in the expressions.
With these conventions, the mean-field energy contribution of a site $i \in A$ can be written as:
\begin{eqnarray} 
\label{eq_Asub_ditmf_en}
\mathcal{E}^{\rm DIT}_{A} &=&
\sum_{\kappa}\bigg\{4T_{\kappa}
        \Big[\eta_{A,\kappa}\phi_{B,\kappa}
        +\phi_{A,\kappa}\eta_{B,\kappa}\Big] 
        \nonumber\\  
	      &+& 4T_{\kappa, 3-\kappa} 
        \Big[\eta_{A,3-\kappa,\kappa}\phi_{B, \kappa} 
        + \phi_{A,\kappa}\eta_{B, 3-\kappa, \kappa} \Big] \bigg\},\nonumber\\
\end{eqnarray}
and  
\begin{eqnarray} 
\label{eq_Asub_nnmf_en}
 \mathcal{E}^{\rm NN} =
\sum_{\kappa} 4\bigg[ \frac{V_{\kappa}}{2} \rho_{A,\kappa}\rho_{B, \kappa} 
+  \frac{V_{\kappa, 3-\kappa}}{2} \rho_{A,\kappa} \rho_{B, 3-\kappa}\bigg],
\end{eqnarray}
due to the DIT processes and NN interactions, respectively.

\begin{table*}[ht]
    \caption{Classification of different quantum phases}
  \begin{ruledtabular}
	  \begin{tabular}{lllccc} 
      \text{Quantum phase} 
          & \hspace{0.1cm}$\rho_i$ 
          & \hspace{0.1cm}$\phi_i$ 
          & \hspace{0.1cm}$\rho_{\kappa}^{\rm SL} := (\rho_{\kappa}^{A}, \rho_{\kappa}^{B})$
		   & \hspace{0.1cm}$\Phi_{\kappa}^{\rm SL} := (\phi_{\kappa}^{A}, \phi_{\kappa}^{\rm B})$ 
          & \hspace{0.1cm} $\eta_{\kappa}^{\rm SL} := (\eta_{\kappa}^{A}, \eta_{\kappa}^{B})$ \\
      \colrule
       Mott insulator (MI) & \text{Integer} & $ = 0$    & $(\rho_{\kappa},\rho_{\kappa})$  & $(0,0)$  &$(0,0)$\\
       Superfluid (SF)     & \text{Real}    & $ \ne 0$  & $(\rho_{\kappa},\rho_{\kappa})$  & $(\phi_{\kappa}, \phi_{\kappa})$  
                           & $(\eta_{\kappa}, \eta_{\kappa})$\\
       Density Wave (DW)   & \text{Integer} & $ = 0$  & $(\rho_{\kappa}^{A},\rho_{\kappa}^{B})$ & $(0,0)$  &$(0,0)$\\
       Super Solid (SS)    & \text{Real}    & $ \ne 0$ & $(\rho_{\kappa}^{A},\rho_{\kappa}^{B})$ & $(\phi_{\kappa}^{A}, \phi_{\kappa}^{B})$
                           & $(\eta_{\kappa}^{A}, \eta_{\kappa}^{B})$\\
       One Body Staggered Superfluid (OSSF) & \text{Real} & $\ne 0$  & $(\rho_{\kappa},\rho_{\kappa})$
					    & $(\phi_{\kappa}, -\phi_{\kappa})$   & $(\eta_{\kappa}, -\eta_{\kappa})$\\
       One Body Staggered Super Solid (OSSS) & \text{Real}  & $ \ne 0$ & $(\rho_{\kappa}^{A},\rho_{\kappa}^{B})$
		                             & $(\phi_{\kappa}^{A}, -\phi_{\kappa}^{B})$  &$(\eta_{\kappa}^{A}, -\eta_{\kappa}^{B})$\\
    \end{tabular}
      \label{table}
  \end{ruledtabular}
\end{table*}
\subsection{Characterization of quantum phases}
\label{sec_cop}
The Hamiltonian in Eqs.(\ref{tbhm}) and (\ref{hamil_terms}) is expected to withstand quantum phases with both uniform and lattice  transnational symmetry broken order parameters. So, we identify different quantum phases based on the combinations of values of different mean-fields and periodicity therein, which stand as order parameters in some cases under the present framework. Particularly we consider the SF order parameter, average density and the order parameter associated with the density inducted transport. 
Since we are working with two-component system, we compute the total SF 
order parameter $\phi_{i} = \phi_{i,1} + \phi_{i,2}$ and  total density $\rho_{i} = \rho_{i,1} + \rho_{i,2}$ of every sites in the lattice to identify the known 
MI and SF phases of two-component BHM \cite{anufriiev_16, bai_20}. In the MI phases of different fillings, the SF order parameter $\phi_i$ is zero, and the system has commensurate integer density $\rho_i \in \mathbb{N}$ on every lattice site. Whereas, for the SF phase 
$\phi_i$ is non zero quantity and $\rho_i\in\mathbb{R}^{+}$.

In the presence of DIT term, the BHM can withstand the OSSF phase, which can be characterized through sign staggering in the phase of SF order parameter of adjacent lattice sites, yielding a checkerboard order in the phase of  $\phi_{i,\kappa}$ for a isotropic 2D square optical lattice. To be explicit in denoting the periodicity in the phase of the SF order parameter, we adapt bipartite lattice description of checkerboard ordering.

Since, spatial profile of $\phi_{i,\kappa}$ has checkerboard sign staggering but uniform $|\phi_{i,\kappa}|$ distribution in the OSSF phase, we can express $(+\phi_{\kappa}, -\phi_{\kappa})$. Additionally, $\rho_{i,\kappa}$ is real and uniform, like in the SF phase. 

With the inclusion of NN density-density interaction term, the model can stabilize DW and SS phases for significant interaction strength. The DW phase, similar to the MI phase, is an insulating phase manifested by zero $\phi_{i,\kappa}$. However, this phase has checkerboard density distribution unlike the uniform MI phase. The SS phase, on the other hand, has the checkerboard profile in spatial distributions of both density and SF order parameters. Following the bipartite lattice notation, we can express the checkerboard pattern in the density distribution of these phases as $\rho_{\kappa}^{\rm SL} = (\rho_{\kappa}^{A}, \rho_{\kappa}^{B})$.
Furthermore, the model in the presence of both DIT and NN interaction terms can exhibit the OSSS phase, which has the checkerboard sign staggering in the phase of $\phi_{i,\kappa}$ along with the checkerboard distribution of $|\phi_{i,\kappa}|$ and $\rho_{i,\kappa}$. The order parameter $\eta_{i,\kappa}$ qualitatively has similar trend as that of $\phi_{i,\kappa}$ in all the mentioned phases, and hence, can be denoted as $\eta_{\kappa}^{\rm SL} = (\eta_{\kappa}^{A}, \eta_{\kappa}^{B})$ in the phases with spatial checkerboard distribution. 

In this work, we obtain the quantum phases in a two component bosonic mixture with symmetric parameters. That is, the strengths of the hoppings, DIT processes, onsite and NN interactions, and chemical potentials of the components are equal, as mentioned earlier. Due to this, the components have identical distribution of the quantities in the miscible phase. 
In contrast, in the immiscible phase the components do not co-exist on same site, but they retain the spatial distribution in the separated regions of the system.  
To summarize, we list the characteristics of different quantum phases in the table \ref{table}. 

We explore the MIT in the strongly interacting regime, particularly in the SF, OSSF, SS and OSSS phases of both components. To quantify the co-existence of the components across the lattice system, we compute the miscibility parameter defined as 
\begin{equation}
 \Lambda = \frac{\frac{1}{L^2}\sum_{i} \rho_{i,1} \rho_{i,2}}
 {(\frac{1}{L^2}\sum_{i} \rho_{i,1})(\frac{1}{L^2}\sum_{i} \rho_{i,2})}.
 \label{ovp}
\end{equation}
Therefore, in the immiscible phase $\Lambda$ is zero, whereas, any finite value of   $\Lambda$ corresponds to partial miscibility of the components. In the regime when the components are completely miscible with density homogeneous distributions,  $\Lambda = 1$. It is to be noted that $\Lambda$ changes from a finite value to zero continuously as the system  parameters are changed to explore the MIT. 
At the verge of this transition, under the considered mean-field framework, we notice replenishment of one of the component at the expense of complete depopulation of the other component.  
This immediately yields $\Lambda= 0$ marking the MIT. In the numerical analysis, this artifact occurs when the initial guess values of the mean-fields are chosen to be homogeneous for both components. But, with spatially asymmetric initial distributions of the mean-fields, both the components remain, and they occupy spatially separated regions in the immiscible phase. Whereas, identical distributions (homogeneous or checkerboard) of the components are retrieved in the miscible phase. 
Such depopulation of a component has also been observed in the context of MI-SF transition under mean-field framework \cite{isacsson_05}.


\section{MIT in the strongly interacting regime}
\label{sec_mitstrong}

MIT has been investigated quite extensively in the multi-component bosonic mixtures confined in continuum potentials. For such systems, the criterion of MIT is predominantly determined by the competition of the intra and inter-component contact interactions \cite{ho_96, ao_98, pu_98, pethick_2002}. This is further explored in shallow optical lattice systems, where both the components are in the weakly interacting SF phase. Like in the continuum systems, criterion for immiscibility is given by  
\begin{equation}
\label{mit_cond_onsite}
U_{12}^2\geq U_{1}U_{2}. 
\end{equation}
On the other hand, in the strongly interacting regime, rich phase space structure arises for the two-component BHM \cite{isacsson_05, anufriiev_16, bai_18}.  
Stemming from the competition of the hopping and contact interactions, both the components can simultaneously be in SF, MI or MI+SF phases. The latter denotes the case where one of the components is in MI phase, whereas the other component is in SF phase. Additionally, the model can withstand counter-flow SF phase and N\'eel ordered anti-ferromagnetic states for $U_{12} > 0$. However, these states are indistinguishable from the MI phase as far as $\rho_i$, $\phi_i$ and individual SF order parameters $\phi_{i,\kappa}$, are concerned \cite{altman_03,kuklov_04,soyler_09,Victor_22}. 

For the MIT, an analogous condition as in the Eq. \eqref{mit_cond_onsite} can be obtained, when both the components are in the MI phase \cite{chen_03}. In earlier works considering 1D optical lattice system with sufficiently number imbalanced mixtures and strong $U_{12}$, metastable non-overlapping emulsion states, exhibiting glassy characteristics have been investigated  \cite{roscilde_07, buonsante_08}. Also, immiscibility on length scales with increasing inter-component repulsion has been reported in Ref \cite{ofir_06}. In the 1D BHM, MIT has been explored when both components are simultaneously MI or SF, and immiscibility condition is obtained in terms of $U_1$, $U_2$ and $U_{12}$ \cite{mishra_07, zhan_14, zhan_14_1}. Furthermore, at half-fillings of the 2D two-component BHM
with symmetric hopping and interaction strengths, immiscible SF and MI phases have been found when the inter-component interaction becomes larger than the repulsive intra-component interaction
\cite{lingua_15}. 
For unequal strengths of hopping processes, such a  transition is noticed between the anti-ferromagnetic N\'eel
state and immiscible SF phase \cite{uttam_10}. The combined effects of contact and light-matter interactions in an optical cavity on the miscibility property of a two-component system has been recently investigated \cite{leon_23}. It is to be noted that in all these studies, the MIT is explored as a consequence of the interplay of on-site interactions, and generically a condition as in Eq. \eqref{mit_cond_onsite} has been considered as the criterion of immiscibility. But, effects of interaction induced processes, predominantly DIT and NN interaction, become relevant in the strongly interacting regime, as mentioned earlier. These are not only known to stabilize new phases and significantly modify the phase diagram, but also expected to influence usual phase transitions of the BHM. We, therefore, explore the effects of these processes on the MIT of several phases. 

In an earlier work \cite{bai_20}, motivated by the experimental realization of the  quantum degenerate mixture of dipolar bosonic atoms \cite{trautmann_18}, we have explored the quantum phases of two-component BHM in the presence of NN interactions. Particularly, we investigated the phases in miscible and immiscible regimes, by considering the $U_{12}<U$ and $U_{12}> U$, respectively, and we noticed the side-by-side density distributions of different immiscible quantum phases.  
In this work, we numerically investigate the MIT of the strongly interacting 
OSSF, SF, SS and OSSS phases in the presence of interaction induced process. 
In particular, we explore the competing roles of the DIT
processes and NN interactions in inducing the MIT of these phases.

\begin{figure}[t]
\includegraphics[width=8.5cm]{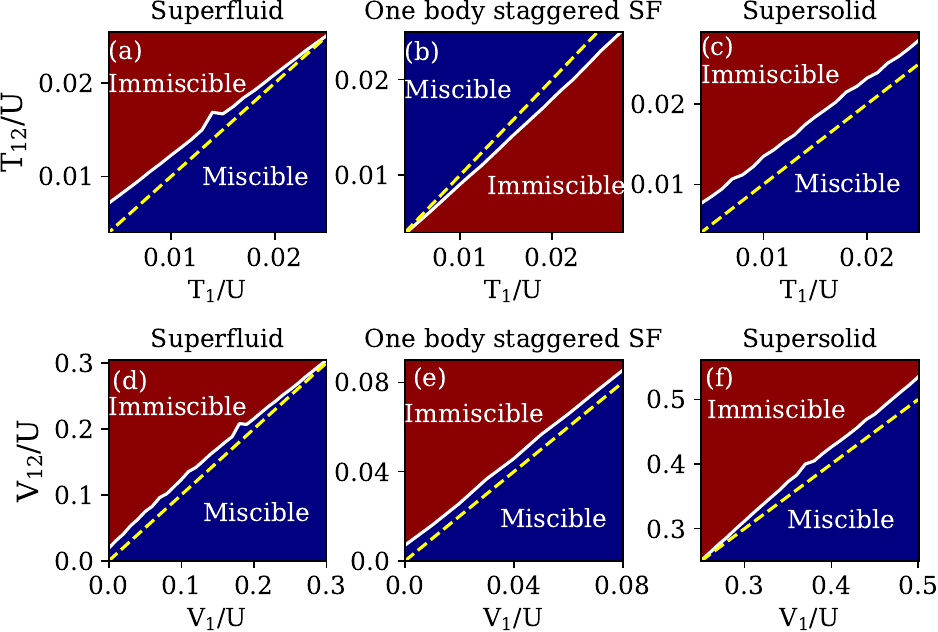}
\caption{MIT of different strongly interacting compressible phases of the model in Eqs. (\ref{tbhm}) and (\ref{hamil_terms}) can be induced through changing the strengths of intra and inter-component DIT and NN interaction processes. (a)-(c) show the phase diagrams for MIT with respect to change of inter and intra-component DIT strengths, $T_1 = T_2$ and $T_{12}$, in the SF, OSSF, and SS phases of both components. (d)-(f) show the similar phase diagrams with respect to change of inter and intra-component NN interaction strengths, $V_1 =V_2$ and $V_{12}$. In our study, we focus in the regime of strong on-site interaction such that the hopping processes are sufficiently weak. Here we scale the system parameters by  $U_1 = U_2 = U_{12} = U$. 
Other system parameters for different figures are: (a) $\mu = 1.9U$, $J = 0.1U$, $V_1 = V_2 = V_{12} = 0$, 
              (b) $\mu = 1.9U$, $J = 0.005U$, $V_1 = V_2 = V_{12} = 0$,
              (c) $\mu = 1.0U$, $J = 0.08U$, $V_1 = V_2 = V_{12} = 0.3U$,
              (d) $\mu = 1.9U$, $J = 0.15U$, $T_1 = T_2 = T_{12}= 0$,
              (e) $\mu = 2.7U$, $J = 0.005U$, $T_1 = T_2 = T_{12} = 0.015U$,
              (f) $\mu = 1.0U$, $J = 0.08U$, $T_1 = T_2 = T_{12} = 0$. The solid white solid line in every panel marks when $\Lambda$ [defined in Eq. \eqref{ovp}] becomes zero in the considered framework. The dashed yellow line mark the identity line to guide reader's eye for the immiscibility condition. 
              }         

\label{fig_mit}
\end{figure}

In Fig. \ref{fig_mit}, we delineate the results on MIT by changing the strengths of intra and inter-component DIT processes and NN 
interactions. To steer the MIT and examine the effects of these processes, we consider $U_{1} = U_{2} = U_{12} = U$. This not only allows us to scale the total Hamiltonian by $U$, but also corresponds to the condition of immiscibility with respect to the onsite contact interaction processes. The strongly interacting regime is when the strengths of the other processes are $\ll U$. In this regime, we carefully choose the other parameters such that the two-component SF, OSSF and SS phases are obtained. For this, it is essential to have knowledge on the phase diagrams of the two-component BHM with the DIT processes and NN interactions. For brevity, such phase diagrams are presented in Figs. \ref{ph_dig}, \ref{ph_dig_vn} and \ref{ph_dig_vn_0pt3}, and explained in detail in Sec. \ref{sec_brod_param}. 
The considered values of the parameters are provided in the caption of Fig. \ref{fig_mit}.

\subsection{MIT with respect to DIT processes}
To begin with, we consider the two-component BHM along with the DIT processes, and take $T_1 = T_2> 0$. Then, for considerable strengths of DIT processes, the model can withstand the two-component OSSF phase, in addition to the two-component MI and SF phases. In Fig. \ref{fig_mit} (a), we depict the MIT when both the components are in the SF phase. We notice that the components remain miscible as long as the inter-component DIT is weaker than the intra-component DIT processes $T_{12}<T_1$. On the other hand, the components become immiscible for $T_{12}>T_1$.  
It is noteworthy that the condition for immiscibility is $T^2_{12} \ge T_1 T_2$, which aligns with Eq. \eqref{mit_cond_onsite}.
Since we take $T_1 = T_2$, the condition simplifies to $T_{12} \ge T_{1}$. For clarity, the threshold line
$T_{12} = T_1$ is indicated with a dashed yellow line in Fig. \ref{fig_mit}. Our numerical analysis is consistent with this immiscibility condition.
However, slight deviations may also arise from the limitations of the mean-field approximation. Nevertheless, the overall behavior remains qualitatively similar to the immiscibility condition expressed in terms of the on-site interaction strengths, as given in Eq.~(\ref{mit_cond_onsite}).
The Hamiltonian in Eqs.~(\ref{tbhm}) and~(\ref{hamil_terms}) has $U(1)$ symmetry for both the components, which yields the energy functional in the even powers of the SF order parameters under the usual Landau framework \cite{iskin_11, bandyopadhyay_19}. In the homogeneous lattice system, the SF phase has uniform density and SF order parameter distributions. Therefore, energy contribution of the DIT processes to the energy functional is positive when the strengths are considered to be positive.
This can also be understood from Eq. \eqref{eq_Asub_ditmf_en} by noting  that the order parameters of the two sub-lattices are same in the uniform SF phase. In addition, the density induced transport order parameters have same sign to that of the SF order parameters, as mentioned in Sec. \ref{ss_mfhamil_energy}. This yields positive mean-field energy contribution due to the DIT processes. 
So, the contribution from the inter-component DIT process becomes energetically costly for the miscible configuration with $T_{12}>T_1$. As a consequence, the components become immiscible, thereby, making the contribution from the inter-component DIT process to diminish. We extend the analysis for the negative values of the DIT strengths, that is, $T_1, T_{12} < 0$, and observe that the system remain immiscible as long as $|T_{12}|<|T_1|$. This is in accordance with the energy cost argument mentioned before.

On the other hand, we observe the components are immiscible when $T_{12}<T_{1}$ in the OSSF phase of the components [see Fig. \ref{fig_mit} (b)]. This is in striking contrast to the trend observed in the SF phase and also to Eq. \eqref{mit_cond_onsite}. 

This seemingly opposite immiscibility condition for positive DIT strengths can be understood through an energy minimization argument. In the OSSF phase, both the SF and DIT order parameters exhibit a checkerboard sign staggering. Within the bipartite lattice convention, these order parameters take the form $\Phi_{\kappa}^{\rm SL} := (\phi_{\kappa}, -\phi_{\kappa})$ and $\eta_{\kappa}^{\rm SL} := (\eta_{\kappa}, -\eta_{\kappa})$, respectively
(see Table \ref{table} and Sec. \ref{sec_cop}). 
As a consequence, the mean-field energy contribution from the DIT processes acquires an overall negative sign [see Eq.~\eqref{eq_Asub_ditmf_en}]. More explicitly, the DIT energy term can be written as:
\begin{eqnarray}
\mathcal{E}^{\rm DIT}&=&
\sum_{\kappa}\bigg\{-4T_{\kappa}
        \Big[\eta_{\kappa}\phi_{\kappa}
        +\phi_{\kappa}\eta_{\kappa}\Big]\nonumber\\
	      &-& 4T_{\kappa, 3-\kappa} 
        \Big[\eta_{3-\kappa,\kappa}\phi_{\kappa} 
        + \phi_{\kappa}\eta_{3-\kappa, \kappa} \Big] \bigg\}.\nonumber
\end{eqnarray}
As discussed earlier, the sign relations $\textrm{sgn}(\eta_{\kappa}) = \textrm{sgn}(\phi_{\kappa})$ and $\textrm{sgn}(\eta_{3-\kappa,\kappa}) = \textrm{sgn}(\phi_{\kappa})$ ensure that this negative prefactor, applied to the positive intra and intercomponent DIT strengths, leads to an energetically favorable miscible configuration when 
$T_{1} < T_{12}$.
Thus, the condition for immiscibility aligns with the case in which the DIT strengths are negative and the components are in the SF phase, as discussed at the end of the previous paragraph.
It is known that positive DIT strength
is essential to withstand the OSSF phase \cite{John_19}. So, we can not explore negative DIT strengths regime for the MIT in this phase.

Next, we examine the immiscibility condition of the components when they are in the SS phase. To obtain the SS phase, we require the NN interaction term in the Hamiltonian as explained in Sec. \ref{sec_model}. We fix the NN interaction strengths $V_1 = V_2 = V_{12}= 0.3 U,$ and obtain the MIT condition with respect to DIT processes in the SS phases of the components, which is shown in Fig~\ref{fig_mit} (c). Here, similar to the case with SF phase, we recover the MIT condition. That is, the components become immiscible when $T_{12}>T_{1}$ and they are miscible when $T_{12}<T_{1}$.
In the SS phase, the components have checkerboard ordering in the density, SF and density induced transport order parameters. However, unlike the OSSF phase, there is no sign staggering in the order parameters (see Table \ref{table}). Therefore, the energy contribution from the intra and inter-component DIT processes is 
positive. Consequently, the components become immiscible when the inter-component DIT strength is larger than the intra-component strengths. The MIT condition remains similar to the SF phase when the DIT strengths are considered negative.

\subsection{MIT with respect to NN interactions}

Here, we examine the MIT by varying the NN interaction strengths. To explore this in the SF and SS phases of the model, the DIT processes can be set to zero, but for the OSSF phase DIT is essential. In Fig \ref{fig_mit} (d)-(f), we depict the MIT when both components are in SF, OSSF, and SS phases, respectively. The DIT strengths are set to zero for Fig. \ref{fig_mit} (d) and (f), but fixed to 
$T_1 = T_{12} = 0.015U$ for \ref{fig_mit} (e). In all these phases, 
similar to the condition in Eq.~\eqref{mit_cond_onsite}, the components are immiscible when $V_{12}>V_1$ and miscible for $V_{12}<V_1$ for positive strengths of the NN interactions.  The energy contribution of the NN interactions is positive for the positive strengths [see Eq. \eqref{eq_Asub_nnmf_en}]. So, the components become immiscible for $V_{12}>V_{1}$, to reduce the energy contribution from the inter-component NN interaction. On the other hand, for negative NN interaction strengths, that is, $V_{1}, V_{12} < 0$, the energy contribution is negative. Consequently, the components remain immiscible $|V_{12}|<|V_{1}|$. 

In Fig \ref{fig_mit} we present the numerical results obtained for a lattice system of size $30 \times 30$.  Increasing the lattice size further does not qualitatively change our results. We illustrate the distributions of density, SF and density induced order parameters of immiscible OSSF, SF, SS phases in the Fig. \ref{den_ossf}, Fig. \ref{den_sf}, and Fig. \ref{den_ss}, respectively.

\begin{figure}[t]
   \includegraphics[width=8.5cm]{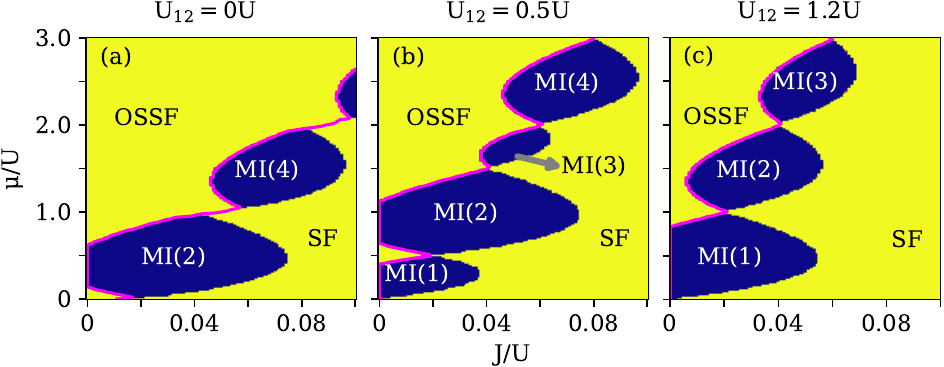}
   \includegraphics[width=8.5cm]{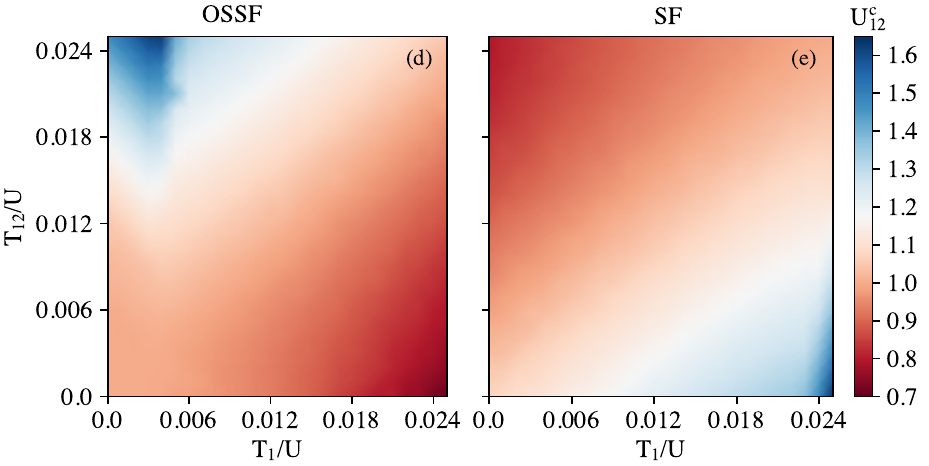}
	\caption{ (a) - (c) Ground state phase diagrams of the two-component BHM in the presence of DIT processes. The phase diagrams are presented for the three values of the inter-component onsite interaction strength $U_{12}$ mentioned at the top of the panels. For all these plots, the strengths of the intra and inter-component DIT processes are fixed as $T_{1} = T_{2} = T_{12} =  0.01 U$. The phase diagrams illustrate parameter domains of the standard two-component SF and MI phases. The parameter domains of the latter correspond to Mott lobes for even and odd fillings. With finite DIT processes the model withstands two-component OSSF phase in the strongly interacting regime (phase boundary is marked by the magenta line). The threshold value of the SF order.  parameter is chosen to be $10^{-6}$ while plotting the phase diagrams in a bi-color scheme (blue and yellow).
    (d) and (e) show the dependence of $U_{12}^{\rm c}$ with respect to $T_1= T_2$ and $T_{12}$ in the OSSF and SF phases, respectively. $U_{12}^{\rm c}$ is the inter-component interaction strength required for immiscibility (implying $\Lambda =0$) when $U_1 = U_2 = U$. The other parameters are fixed as $\mu = 1.9 U$, $ J = 0.005 U$ for the OSSF phase and $\mu = 1.9 U$, $J = 0.1 U$ for SF phase.}
     \label{ph_dig}
\end{figure}

\section{Broadening the parameter regime}
\label{sec_brod_param}
In this section, we extend our analysis on the MIT of the two-component system with DIT processes and NN interactions. We aim to probe the effects of these processes on the immiscibility condition in terms of the on-site interaction strengths [see Eq. \eqref{mit_cond_onsite}]. To explore this in the strongly interacting phases of the model, the knowledge of the phase diagram is essential. For single component bosons in 1D and 2D optical lattices with DIT process, the mean-field phase diagrams have been reported \cite{John_19, suthar_20_2, suthar_22}. But, the phase diagram for the two component BHM with DIT processes
has not been reported yet. Therefore, we obtain the mean-field phase diagram of a two-component BHM with DIT processes. We first present the phase diagrams in Figs. \ref{ph_dig} (a) - (c) for different values of inter-component interaction strengths for the case when the NN interactions are absent. Later, similar phase diagrams are presented for finite NN interaction strengths in Figs. \ref{ph_dig_vn} (a) - (c).

\subsection{Phase diagram of two component BHM with DIT}
\label{sec_phdiag_bhm_dit}

We consider a 2D square lattice of size $10 \times 10$ with periodic boundary conditions to numerically obtain the phase diagrams. 
For this, we employ the framework explained in Secs. \ref{sec_mft} and \ref{sec_num}, and evaluate the order-parameters to characterize the various quantum phases (see Sec. \ref{sec_cop} and Table \ref{table}) of the model. The scale is fixed as $U_{1} = U_{2}=  U$. Then, the phase diagrams are obtained for three different values of the inter-component interaction strengths, $U_{12}$, which are presented in Fig.~\ref{ph_dig} (a) - (c). In the absence of DIT processes, the mean-field phase diagrams of a two-component BHM have the parameter domains for two-component SF and MI phases \cite{anufriiev_16, bai_20}. The parameter domains of the MI phase of different fillings form usual Mott lobes which are marked as MI($\rho_i$), where $\rho_{i} = \rho_{i,1} + \rho_{i,2}$. On the other hand, finite DIT processes destabilize the insulating MI phase towards OSSF phase in the strongly interacting regime. For the phase diagram simulations, we consider $T_{1} = T_{2} = T_{12} = 0.01 U$. In the absence of inter-component interaction, $U_{12} = 0$, we obtain Mott-lobes of only even integer $\rho_{i}$ [see Fig.~\ref{ph_dig} (a)], which we refer to as even MI lobes. These also correspond to MI lobes of integer filling. In these two-component MI phases, both the components have uniform commensurate integer occupancy and zero SF order parameter.
With increasing density the hopping of the particles in the system is enhanced due to the DIT processes. Thereby, the parameter domain of the two-component OSSF phase gets larger, a characteristic also present in the single-component case \cite{John_19}. The phase boundary of the OSSF phase is represented by magenta line in Fig.~\ref{ph_dig} (a) - (c). 

Finite inter-component onsite interaction introduces the parameter domains for MI phases of odd integer $\rho_i$, which form the odd MI lobes [see Fig.~\ref{ph_dig} (b) and (c)]. In these MI phases, total density distribution is uniform and integer, but the density distribution of individual component is random. The SF order parameters are zero and occupancy of each component at any site is integer. Consequently, these are incompressible phases with half-integer filling, for example, the MI(1) and MI(3) correspond to two-component filling $1/2$ and $3/2$, respectively.
These MI phases are reminiscence of the counter-flow SF phases, which are indistinguishable from the insulating Mott phases in the present mean-field framework. In a recent experiment the counter-flow SF phase has been observed \cite{zheng_24}. 
The odd MI lobes enhances with the increase of $U_{12}$, whereas, the size of even lobes remain the same until the immiscibility condition is satisfied. However, these lobes are pushed towards larger values of $\mu$ due to the growth of odd lobes. These characteristics are evident from the comparison of Figs.~\ref{ph_dig} (a) - (c). An important feature of the phase diagram is that the parameter domain of the two-component OSSF phase reduces with the increase of $U_{12}$. The stronger inter-component interaction stabilizes the insulating states against the OSSF phase. 

Since, we extend our analysis on the MIT of different phases to the negative values of the DIT strengths (Sec. \ref{sec_mitstrong}), here we remark the effects of negative DIT strengths on the phase diagram. The OSSF phase only stabilized for the positive DIT strengths.  For negative DIT strengths, the DIT processes play similar role to that of the hopping processes, which is also evident from Eq. \eqref{tbhm}. We can think of the effective strengths of the hopping processes get enhanced for negative values of the DIT strengths. This effect is stronger for larger occupancy, hence, for larger values of $\mu$.  Consequently, the SF phase favoured for negative DIT strengths and usual MI lobes shrink in size. For significantly large DIT strengths, MI lobes disappear and the phase diagram in the positive $J/U - \mu/U$ plane trivially has SF phase.

\subsection{Influence of DIT processes on the MIT of OSSF and SF phases}
\label{sec_influ_dit_bhm}
Here, we examine the effects of the DIT processes on the usual MIT of OSSF and SF phases, when the NN interactions are absent.
For this, we numerically obtain the immiscibility condition in terms of inter-component onsite interaction strength as the intra and inter-component DIT strengths, $T_1 = T_2$ and $T_{12}$, are varied. That is, we investigate the MIT criterion as in Eq. \eqref{mit_cond_onsite} in the presence of DIT processes. The inter-component interaction strengths are fixed at $U_{1} = U_{2} = U$. Then, we find the required strength of $U_{12}$ for the components to be immiscible ($\Lambda$ becomes zero), which we denote as $U_{12}^{\rm c}$. We illustrate the variation of $U_{12}^{\rm c}$ in the heatmaps shown in Fig.~\ref{ph_dig}(d) and Fig.~\ref{ph_dig}(e) for the OSSF and SF phases, respectively. To explore the condition in the OSSF phase [see Fig. ~\ref{ph_dig}(d)], we choose the chemical potential $\mu = 1.9 U$ and hopping strength $J= 0.005U$. 
For $T_{1} = T_{2} = T_{12} = 0$, we recover the known immiscibility condition $U_{12}^{\rm c} = U$. We observe that $U_{12}^{\rm c}$ is decreased with the increase in $T_1$ for fixed $T_{12}$, while it increases with the increase in $T_{12}$ for fixed $T_1$. On the other hand, in the SF phase, the immiscibility condition gets modified in opposite way [see Fig. ~\ref{ph_dig}(e)]. The corresponding heatmap is obtained for $\mu = 1.9 U$ and $J = 0.1U$. We observe that $U_{12}^{\rm c}$ is decreased (increased) with the increase (decrease) in $T_{12}$ ($T_1$). Additionally, in both the cases $U_{12}^{\rm c} = U$ for $T_{1} = T_{12}$. These observations demonstrate that the MIT gets significantly influenced by the DIT processes. Furthermore, intra-component DIT processes favour immiscibility by lowering the value $U_{12}^{\rm c}$ in the OSSF phase, whereas they favour miscibility in the SF phase. 
This opposite trend of $U_{12}^{\rm c}$ in Fig.~\ref{ph_dig}(d) and Fig.~\ref{ph_dig}(e) is consistent with the observations made in Fig.~\ref{fig_mit} and the arguments presented in Sec.\ref{sec_mitstrong}.

Note that, so far we have considered fixed hopping strengths. This allows us to explore the MIT in the targeted quantum phases. But, we also extend our study for other values of the hopping strengths in Appendix. \ref{append_influ_dit}. 
With the change in hopping strengths, the model is expected to show transitions between the OSSF, MI and SF phases. Therefore, we can probe the influence of DIT processes on the MIT of these phases.


\begin{figure}[t]
   \includegraphics[width=8.5cm]{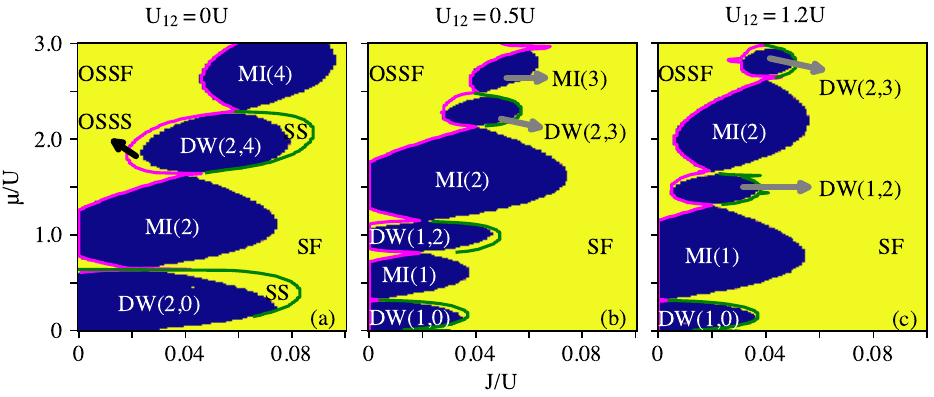}
    \caption{(a) - (c) Ground state phase diagrams of the two component BHM in the presence of DIT processes and NN interactions. The phase diagrams are presented for the three values of the inter-component onsite interaction strength $U_{12}$ mentioned at the top of the panels. For all these plots, $T_{1} = T_{2} = T_{12} =  0.01 U$ and $V_{1} = V_{2}= V_{12} = 0.08U$. The phase diagrams illustrate lobes of two-component MI and DW phases with different fillings. 
    NN interaction stabilizes parameter domains of the SS phase in between the the DW lobes and SF phase. Additionally, finite DIT processes and NN interactions are responsible for stabilizing OSSF and OSSS phases. The phase boundary of the SS phase is marked by the solid green line. The solid magenta line correspond to the phase boundary of the OSSF phase. The threshold value of the SF order.  parameter is chosen to be $10^{-6}$ while plotting the phase diagrams in a bi-color scheme (blue and yellow).
    }
     \label{ph_dig_vn}
\end{figure}

\subsection{Phase diagram of two component BHM with DIT and NN interactions}
\label{sec_phdiag_bhm_dit_nn}
We discuss the phase diagrams of the BHM in the presence of both DIT processes and NN interactions, which are presented in Fig. \ref{ph_dig_vn} (a) - (c) for three different values of $U_{12}$. Similar to the previous case, the scale is fixed by $U_1 = U_2 = U$.
We consider $T_1 = T_2 = T_{12} = 0.01U$ and $V_{1} = V_{2} = V_{12} = 0.08U$ for the phase diagrams. For this weak value of the NN interaction strengths, the phase diagram remains qualitatively similar to that of the Fig. \ref{ph_dig}, that is, the parameter domains of the compressible phases surround the lobes of incompressible phases.
However, the considered NN interaction strength is sufficient to withstand the quantum phases with checkerboard ordering in the density distribution, thereby stabilizing the DW and SS phases. Then, the phase diagram has parameter domains for the two-component DW and SS phases in addition to the MI and SF phases. For $U_{12} = 0$, we obtain the lobes of DW(2,0), MI(2), DW(2,4), MI(4), and so on, with increasing $\mu$ [see  Fig.~\ref{ph_dig_vn} (a)]. Here, we have adapted the bipartite lattice notation (discussed in Sec.\ref{sec_cop}) with belonging to the two sub-lattices to mark the DW lobes. Note that these lobes are of incompressible phases with two-component filling equals to $1/2$, $1$, $3/2$, and $2$. The lobes for integer and half-integer filling are for MI and DW phases, respectively. The parameter domains of the SS phase envelope the DW(2,0) and DW(2,4) lobes, and its phase boundary is marked by the solid green line. 
Similar to Fig. \ref{ph_dig}, the phase diagram has the parameter domain of the OSSF phase to the left of the lobes of the incompressible phases. In this case, we additionally obtain parameter domains of the OSSS phase which is stabilized by both the DIT processes and NN interactions. 

With the increase of $U_{12}$, like in the previous case, Mott lobes with half-integer filling [for example, MI(1) and MI(3)] emerge [see Fig. \ref{ph_dig_vn}(b)]. Additionally, DW phases with quarter filling, such as, DW(1,0), DW(1,2), DW(2,3), etc., get stabilized. Note that these DW phases have two-component filling equals to $1/4$, $3/4$ and $5/4$, respectively. Like in the previous case the SS and OSSS phases have parameter domains enveloping the the DW phases. But, it is important to note that the parameter regime of SS and OSSS phases shrink with increasing $U_{12}$, which is evident from the comparison between Figs. \ref{ph_dig_vn}(a)-(c). This trend is consistent as increasing inter-component onsite interaction results in stabilizing the insulating phases against the compressible SS and OSSS phases.  For $U_{12} = 1.2 U$, the parameter domains of these phases are marginally present [see Fig. \ref{ph_dig_vn}(c)]. Therefore, it is essential to consider larger NN interaction strengths to analyze the MIT of the SS phase. In this case, the phase diagram has lobes of DW(1,0), MI(1), DW(1,2), MI(2), and DW(2,3), corresponding to incompressible phases with two-component filling equals to $1/4$, $1/2$, $3/4$, $1$ and $5/4$, respectively.

\begin{figure}[t]
  \includegraphics[width=4.5cm]{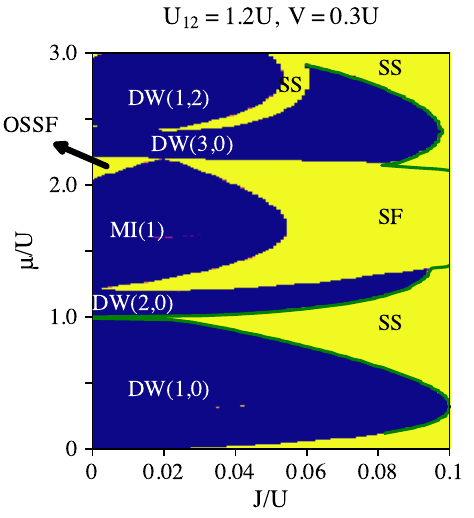}
    \caption{Ground state phase diagram of the the two component BHM with $V_{1} = V_{2} = V_{12} = 0.3U$ and $T_{1} = T_{2} = T_{12} = 0.01 U$. Here we present the diagram for $U_{12} = 1.2 U$, which can be contrasted with the phase diagram of Fig. \ref{ph_dig_vn}(c). Enhanced parameter domain of the SS phase is obtained. The threshold value of the SF order.  parameter is chosen to be $10^{-6}$ while plotting the phase diagrams in a bi-color scheme (blue and yellow).} 
     \label{ph_dig_vn_0pt3}
\end{figure}

We now explore the phase diagram of the model with $V_{1} = V_{2} = V_{12} = 0.3U$ but keeping DIT strengths as previous. For such strong NN interaction strengths, we obtain larger parameter domain of the SS phase. Since with increasing $U_{12}$ the parameter region of the SS phase shrinks, an observation mentioned in the previous paragraph, here we present the phase diagram only for $U_{12} = 1.2 U$ for brevity. In contrast to the case considered earlier, it is important to note that the stronger NN interactions significantly change the phase diagram, which is shown in Fig. \ref{ph_dig_vn_0pt3}. We notice enhanced usual lobes of DW(1,0), MI(1) and DW(1,2), in comparison to the Fig. \ref{ph_dig_vn}(c). Additionally, deformed parameter domains of DW(2,0) and DW(3,0) are obtained next to MI(1) and DW(1,2). Note MI(1) and DW(2,0) phases have same two-component filling equals to $1/2$. Similarly, filling of the DW(1,2) and DW(3,0) phases is $3/4$. Therefore, the strong NN interaction results in
additional DW phases. Interestingly, the transitions from MI(1) to DW(2,0) and DW(1,2) to DW(3,0) are intervened by the the parameter domains SF and SS phases, respectively.
Similar to previous case, the SS phase appears next to the DW phases. However, it is important to note that we obtain significantly large parameter domain of the SS phase in contrast to the Fig. \ref{ph_dig_vn}(c). Therefore, we can examine the influence of the DIT processes on the MIT in the two-component SS phase by making appropriate choice of the parameters.  
Furthermore, we also observe a small domain of the OSSF phase next to the left side of the MI(1) lobe.
%

\begin{figure}[t]
   \includegraphics[width=8.7cm]{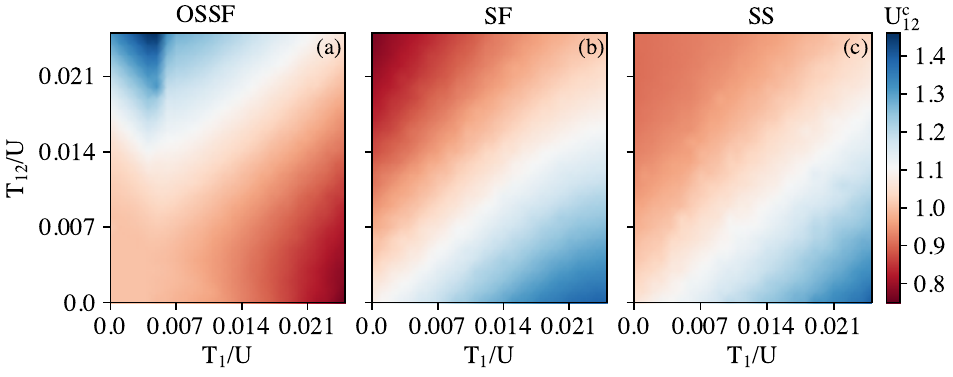}
   \caption{(a)-(c) show the dependence of $U_{12}^{\rm c}$ with respect to $T_1= T_2$ and $T_{12}$ in the OSSF, SF and SS phases, respectively, when the NN interactions are present. $U_{12}^{\rm c}$ is the inter-component interaction strength required for immiscibility (implying $\Lambda = 0$) when $U_1 = U_2 = U$.
   The NN interaction strengths are fixed to $V_1 = V_2 = V_{12} = 0.08U$ for the OSSF and SF phases, while $V_1 = V_2 = V_{12} = 0.3U$ for the SS phase. Other parameters are considered as $\mu = 2.5 U$, $J = 0.005 U$ for (a), $\mu = 1.9 U$, $J = 0.1 U$ for (b), and $\mu = 1.0 U$, $J = 0.08 U$ for (c). }
  \label{influ_dit}
\end{figure}
%


\subsection{Influence of DIT processes on the MIT of OSSF, SF and SS phases}
\label{sec_influ_dit_nn_bhm}

We now explore the effects of the DIT processes on the MIT of the OSSF, SF, and SS phases when the NN interactions are considered in the model. We consider the $V_{1} = V_{2} = V_{12} = 0.08U$ while examining the immiscibility condition for the OSSF and SF phases, whereas it is fixed to 
$V_{1} = V_{2} = V_{12} = 0.3U$ for the SS phase. Like earlier, we find the the required inter-component interaction strength, $U_{12}^{\rm c}$, for the immiscibility ($\Lambda = 0$) as $T_1 = T_2$ and $T_{12}$ are varied. The corresponding heatmaps are depicted in Fig. \ref{influ_dit}. We consider $\mu = 2.5 U$ and $J = 0.005 U$ for the OSSF phase
[see Fig. \ref{influ_dit}(a)] and $\mu = 1.9 U$ and $J = 0.1 U$ for the SF phase [see Fig. \ref{influ_dit}(b)]. We observe qualitatively similar behaviour of $U_{12}^{\rm c}$ as in Figs. \ref{ph_dig}(d) and (e). Thus, the immiscibility is favoured by the intra-component DIT  processes in the OSSF phase. On the other hand, the dependence of $U_{12}^{\rm c}$ in the SS phase [see Fig. \ref{influ_dit}(c)] is similar as in the SF phase. That is, $U_{12}^{\rm c}$ decreases as we increase $T_{12}$ keeping $T_{1}$ fixed, whereas it increases as we increase $T_{1}$ 
keeping $T_{12}$ fixed. Thus, in the SF and SS phases, the immiscibility is favoured by inter-component DIT processes. Guided by the phase diagram presented in Fig. \ref{ph_dig_vn_0pt3}, we consider $\mu = 1.0 U$ and $J = 0.08 U$ to explore the effects of the DIT processes on MIT in the SS phase. This ensures that the two-component SS phase is always retained as the $U_{12}$ is varied to obtain the $U_{12}^{\rm c}$.

\section{Summary and discussion}
\label{sec_summary}
In summary, we have theoretically investigated the MIT of a two-component bosonic mixture in the strongly interacting regime. Our study reveals that the MIT can be steered by controlling the DIT processes and NN interactions, when the mixture is tuned close to the immiscibility with respect to onsite interactions. We particularly study the MIT in the two-component OSSF, SF and SS phases, which are compressible quantum phases of the BHM model with DIT and NN interaction terms. Our investigation suggests $T_{12}>\sqrt{T_{1}T_{2}}$ as the immiscibility condition in the SF and SS phases, which is similar to the immiscibility condition in terms of onsite interaction strengths. In striking contrast, we obtain $T_{12}<\sqrt{T_{1}T_{2}}$ as the immiscibility condition for the OSSF phase. This characteristic arises due to the phase staggering of the OSSF phase. On the other hand, in all these phases the immiscibility condition is found to be $V_{12}>\sqrt{V_1 V_2}$ when the MIT is driven with respect to the NN interactions. 

Next, we have demonstrated that the DIT can significantly influence the MIT of the quantum phases in strongly interacting regime. For this we have analyzed the required inter-component interaction strength for immiscibility of the two components, $U_{12}^{\rm c}$. Our numerical analysis reveals that $U_{12}^{\rm c}$ is considerably lowered or enhanced in the presence of DIT processes. Consistent with the above conclusion, we find that intra-component DIT processes favour immiscibility by lowering the value of $U_{12}^{\rm c}$ in the OSSF phase. Whereas, miscibility is favoured by the inter-component DIT process, since $U_{12}^{\rm c}$ is increased with the increase of $T_{12}$ for fixed $T_1$. We obtain the opposite dependence of $U_{12}^{\rm c}$ on the intra and inter-component DIT processes in the SF and SS phases. Here, the value of $U_{12}^{\rm c}$ is significantly lowered with the increase in $T_{12}$, implying the inter-component DIT processes favour immiscibility of the components. We have first obtained this dependence in the absence of NN  interactions, which we latter corroborate to the case for finite NN interaction strengths. We obtain qualitatively analogous trend in both these cases. Furthermore, we present mean-field phase diagrams of the two-component BHM with the DIT and NN interaction terms.

Interaction induced processes are known to influence several equilibrium and dynamical properties of optical lattice systems in the strongly interacting regime. In this regime, the DIT can arise from strong onsite interaction or due to long-range interaction between the particles. Whereas, the NN interaction becomes relevant in long-range interacting systems. Therefore, the strengths of these processes can be tuned by changing the relevant interaction strength. In an optical lattice system with neutral atoms, the DIT strength can be varied by changing the scattering length for onsite interaction through Feshbach resonances \cite{chin_10}. In this case, the DIT can be tuned to the same order of usual hopping process~\cite{Meinert_13, Jurgensen_14}. On the other hand, atoms or molecules having large dipole moment can interact additionally through long-range dipole-dipole interaction \cite{sowinski_12, baier_16, lukas_17}. This, not only contributes to the onsite interaction, but also gives rise to DIT processes and NN interactions \cite{sowinski_12, baier_16}. The strength of dipole-dipole interaction between the constituent particles, can be controlled by tilting the polarization axis of the dipoles with respect to lattice plane \cite{baier_16, bandyopadhyay_19} or by rapidly rotating the polarization axis \cite{giovanazzi_02, tang_18}. Consequently, strengths of the DIT process and NN interaction can be tuned. Recently, the DIT has also been observed in a minimal set of three Rydberg atoms using dipolar exchange interactions, where DIT occurs from the second order virtual hopping process \cite{Lienhard_20}.
Effects of the DIT process on the usual MI-SF transition have been demonstrated in optical lattice system with neutral \cite{mark_11,Jurgensen_14} and magnetic atoms \cite{baier_16}. Additionally, influence of the DIT on quantum criticality of an effective Ising model has been studied in an optical lattice system with neutral atoms \cite{Meinert_13}. 
To understand the relative strengths of different processes in the presence of an optical lattice potential, consider a lattice with lattice constant \(a \approx 500\,\text{nm}\) and depth \(V_0 = 10 E_R\), where \(E_R\) is the recoil energy. For dipolar atoms with a dipole moment \(\mu \approx 10\mu_{\rm B}\) and an \(s\)-wave scattering length \(a_s = 100 a_0\), the on-site interaction is approximately \(U \sim 1E_R\), and the tunneling amplitude is \(J \sim 0.01E_R\). The nearest-neighbor (NN) interaction strength is of the order \(10^{-3} \gamma E_R\), while the dipolar-induced tunneling (DIT) strength is of the order \(10^{-3} \alpha E_R\), where \(\alpha\) can be tuned via the scattering length~\cite{chin_10}. A detailed estimation of these parameters is provided in Appendix~\ref{app_d}.

Our study concludes that the MIT of a two-component system can be influenced significantly by the DIT processes. Due to the high degree of control on tuning the strengths of the DIT and NN interactions, as discussed above, our findings can be probed in future experiments on optical lattice loaded with mixture of ultracold atoms.

%
\acknowledgments{
We gratefully acknowledge useful discussions with Philipp Hauke.
R.B.\ acknowledges International Institute of Information Technology Hyderabad for kind hospitality during the progress of this work. S.B.\ acknowledges funding by the 
European Union under NextGenerationEU Prot. n. 2022ATM8FY (CUP: E53D23002240006), European Research Council (ERC) under the European Union’s Horizon 2020 research and 
innovation programme (grant agreement No 804305), Provincia Autonoma di Trento, Q$@$TN, the joint lab between University of Trento, FBK-Fondazione Bruno Kessler, 
INFN-National Institute for Nuclear Physics and CNR-National Research Council. Views and opinions expressed are however those of the author(s) only and do not necessarily 
reflect those of the European Union or European Commission. Neither the European Union nor the granting authority can be held responsible for them. 
S.B.\ acknowledges CINECA for the use of HPC resources under ISCRA-C projects ISSYK-2 (HP10CP8XXF) and DISYK (HP10CGNZG9). 
R.B. acknowledges the support of the Deutsche Forschungsgemeinschaft (DFG, German Research Foundation) under Germany’s Excellence Strategy – EXC-2123 Quantum- Frontiers – 390837967.}
\appendix
\section{Details of the Hamiltonian and mean-field methodology}
\label{append1}
In this appendix, we provide further details of the Hamiltonian terms in Eqs~(\ref{tbhm}) and~(\ref{hamil_terms}), and discuss the phases supported by the model. Next, we describe the mean-field decoupling of the Hamiltonian terms and present the Gutzwiller ansatz. Finally, we outline the numerical methodology employed in our work.

In the strongly interacting regime, $J/U \ll 1$, the single component 
BHM exhibits a quantum phase transition between the MI and SF phases stemming 
from the competition of the hopping and onsite interaction processes.  
The system exhibits insulating MI phases of different fillings determined 
by $\mu$ in the strongly interacting limit, $J/U \rightarrow 0$.
While in the weakly interacting regime $J/U \gg 1$ it is in the SF phase. 
The two-component version of the BHM has also been studied in earlier works, 
and known to host additional paired and anti paired 
phases \cite{kuklov_03, kuklov_04}.
In the strongly interacting regime, the phase diagram can have 
parameter domains of nontrivial MI lobes with half-integer total filling 
alongside the SF and usual two-component MI lobes of integer total filling
\cite{isacsson_05, bai_20}. 
Additionally, in this regime the two-component BHM for different fillings 
can be mapped to a plethora of spin Hamiltonians \cite{altman_03, kuklov_03}, 
and thus the two-component mixtures of ultracold atoms in the optical lattices 
have potential applications as quantum simulator of various spin 
models \cite{morera_19}.  

The second term in Eqs. (\ref{tbhm}) and (\ref{hamil_terms}), $\hat{H}^{\rm DIT}$, 
is the Hamiltonian part representing DIT processes of the two-components 
\cite{Jurgensen_12, Luhmann_12, sowinski_12, maik_13}. 
Here, $T_{\kappa}$ and $T_{12} = T_{21}$ correspond to the intra and inter-component 
DIT strengths, respectively. 
For large onsite interaction, the DIT can stem from interaction induced band 
mixing \cite{dutta_11, Luhmann_12, maik_13}, and the process can be 
sufficiently strong for other interactions, such as, dipole-dipole 
interaction \cite{sowinski_12}. 
The inclusion of DIT term in the BHM modifies the phase 
diagram significantly in the strongly interacting regime and for large $\mu$. 
The effect of the latter is intuitively expected as the hopping process is 
enhanced with increasing filling resulting destruction of certain insulating 
phases in favor of superfluidity. Additionally, the system is known to exhibit 
a superfluid phase with uniform density but sign staggering in the phase of 
the SF order parameter, which is referred to as one body staggered superfluid 
(OSSF) phase \cite{John_19}. This staggered phase is a mean-field incarnation
of the novel twisted SF phase having time-reversal symmetry broken 
complex SF order parameter, and this phase is known to be better stabilized in
the multi-component system of bosons \cite{panahi_12}. 

The third term in Eqs. (\ref{tbhm}) and (\ref{hamil_terms}), $\hat{H}^{\rm NN}$, 
stands for the density-density interaction between the particles in the NN sites. 
For a system of strong enough contact interaction this 
process can be relevant, but it is weaker in strength than the DIT 
process \cite{mazzarella_06}. On the other hand, this interaction can be strong 
for atoms or molecules with large dipole moments. Then, the particles in 
different sites can interact through long-range dipole-dipole interaction of 
which the range can be limited to the NN sites to capture the effects of the 
strongest off-site interaction \cite{baier_16, bandyopadhyay_19, bandyopadhyay_22}. 
Such a model with NN density-density interaction 
is traditionally been termed as extended BHM (eBHM) \cite{kuhner_00, sengupta_05, 
scarola_05, mazzarella_06, ng_08, iskin_11, uwe_11, suthar_20_1}, and it is 
experimentally studied with magnetic Er atoms in a 3D cubic and
multi-layers of 2D optical lattice in the context of MI-SF transition 
\cite{baier_16}. In the equations, $V_{\kappa}$ and $V_{12} = V_{21}$ are the 
strengths of intra and inter-component NN interactions, and potential 
realization of the two-component version of the eBHM can be made with the 
mixture of dipolar atoms or molecules \cite{trautmann_18}.
In the presence of NN interaction, the eBHM can withstand quantum phases with 
broken lattice transnational symmetry, such as, DW and SS.
In a square lattice, these phases exhibit a checkerboard pattern in density and SF order parameter. However, phases with structural order extending beyond a single unit cell can be stabilized by models incorporating interactions beyond nearest neighbors—such as next-nearest, next-next-nearest, or dipolar couplings. These extended interactions introduce competing length scales, which may lead to frustration and metastable states, thereby enriching the phase diagram~\cite{Menotti_07, Trefzger_08, Sansone_10}

With the inclusion of the DIT process, the eBHM can furthermore stabilize one 
body staggered supersolid (OSSS) phase, which has the density modulation alongside 
the sign staggering in the phase of the SF order parameter. Note that the Hamiltonian 
in Eqs. (\ref{tbhm}) and (\ref{hamil_terms}), has $U(1)\otimes U(1)$ symmetry due to the 
conservation of total particle number of each component. Additionally, for symmetric choice
of intra-component parameters, the Hamiltonian has additional $Z_{2}$ exchange symmetry. 
The mean-field framework, which we employ to solve the model, explicitly breaks the $U(1)\otimes U(1)$ symmetry, and thereby introduce the SF order parameters into the Hamiltonian depicting the number fluctuations.

\subsection{Mean-field decoupling scheme and mean-field Hamiltonian}
\label{sec_mft}
The different terms of the Hamiltonian in Eq. (2), to leading order in the fluctuations, can be written as

\begin{eqnarray}
	&\hat{b}_{i, \kappa}^{\dagger }\hat{b}_{j,\kappa}&\approx\hat{b}_{i, \kappa}^{\dagger}
  \langle \hat{b}_{j, \kappa}\rangle + \langle \hat{b}_{i, \kappa}^{\dagger} 
  \rangle\hat{b}_{j,\kappa} - \langle \hat{b}_{i,\kappa}^{\dagger}\rangle  
	\langle \hat{b}_{j,\kappa}\rangle, \nonumber\\   &\hat{b}_{i,\kappa}^{\dagger}\hat{n}_{i,\kappa} \hat{b}_{j, \kappa}
   &\approx \hat{b}_{i,\kappa}^{\dagger}\hat{n}_{i,\kappa} \langle\hat{b}_{j,\kappa}\rangle 
    + \langle\hat{b}_{i, \kappa}^{\dagger}\hat{n}_{i, \kappa} \rangle \hat{b}_{j,\kappa}
     \nonumber\\
    &&- \langle\hat{b}_{i, \kappa}^{\dagger}\hat{n}_{i, \kappa} \rangle 
	\langle\hat{b}_{j, \kappa} \rangle, \nonumber \\ 
  &\hat{b}_{i, \kappa}^{\dagger} \hat{n}_{j, \kappa}\hat{b}_{j, \kappa}
   &\approx \hat{b}_{i,\kappa}^{\dagger}\langle \hat{n}_{j, \kappa}\hat{b}_{j,\kappa}\rangle
    + \langle \hat{b}_{i,\kappa}^{\dagger} \rangle \hat{n}_{j,\kappa}\hat{b}_{j,\kappa}
    \nonumber\\
    &&- \langle \hat{b}_{i,\kappa}^{\dagger} \rangle \langle 
	\hat{n}_{j,\kappa}\hat{b}_{j,\kappa}\rangle, \nonumber \\
  &\hat{b}_{i,\kappa}^{\dagger}\hat{n}_{i,3-\kappa} \hat{b}_{j,\kappa}
	&\approx \hat{b}_{i, \kappa}^{\dagger}\hat{n}_{i, 3-\kappa} \langle\hat{b}_{j, \kappa}\rangle 
	+ \langle\hat{b}_{i, \kappa}^{\dagger}\hat{n}_{i, 3-\kappa} \rangle \hat{b}_{j, \kappa}
	\nonumber\\
    &&- \langle\hat{b}_{i, \kappa}^{\dagger}\hat{n}_{i, 3-\kappa}\rangle 
	\langle\hat{b}_{j, \kappa} \rangle, \nonumber \\ 
	&\hat{b}_{i, \kappa}^{\dagger} \hat{n}_{j, 3-\kappa}\hat{b}_{j, \kappa}
	&\approx \hat{b}_{i, \kappa}^{\dagger}\langle \hat{n}_{j, 3-\kappa}\hat{b}_{j, \kappa}\rangle
	+ \langle \hat{b}_{i, \kappa}^{\dagger} \rangle \hat{n}_{j, 3-\kappa}\hat{b}_{j, \kappa}
    \nonumber\\
    &&- \langle \hat{b}_{i, \kappa}^{\dagger} \rangle \langle 
	\hat{n}_{j, 3-\kappa}\hat{b}_{j, \kappa}\rangle, \nonumber\\
	&\hat{n}_{i, \kappa}\hat{n}_{j, \kappa}&\approx\hat{n}_{i, \kappa}
  \langle \hat{n}_{j, \kappa}\rangle + \langle \hat{n}_{i, \kappa} 
  \rangle\hat{n}_{j, \kappa} - \langle \hat{n}_{i, \kappa}\rangle  
	\langle \hat{n}_{j, \kappa}\rangle, \nonumber\\ 
	&\hat{n}_{i, \kappa}\hat{n}_{j, 3 -\kappa}&\approx\hat{n}_{i, \kappa}
	\langle \hat{n}_{j, 3-\kappa}\rangle + \langle \hat{n}_{i, \kappa}
	\rangle\hat{n}_{j, 3 - \kappa} 
    \nonumber \\
    &&- \langle \hat{n}_{i, \kappa}\rangle  
	\langle \hat{n}_{j, 3-\kappa}\rangle,
    \label{mean-field-terms}
\end{eqnarray}
where the mean-field $\phi_{i,\kappa} = \langle \hat{b}_{i,\kappa}\rangle$ is the SF order 
parameter,  $\rho_{i,\kappa} = \langle \hat{n}_{i,\kappa} \rangle$ is the local density, 
$\eta_{i,\kappa} = \langle\hat{n}_{i,\kappa}\hat{b}_{i,\kappa}\rangle$,
$\eta_{i, 12} = \langle\hat{n}_{i,1}\hat{b}_{i,2}\rangle$, and 
$\eta_{i, 21} = \langle\hat{n}_{i,2}\hat{b}_{i,1}\rangle$
are the order-parameters for density-dependent transport properties \cite{John_19}. 

We now explicitly present the single-site mean-field for a 2D square lattice system. In 2D, a lattice index can be written as $i\equiv (p, q)$. The coordination number of a square lattice is $4$, and the NN sites can have index $j\equiv (p^\prime, q^\prime)\in\{(p+1, q), (p-1, q), (p, q+1), (p, q-1) \}$. By adapting a notation $\bar{x}_{p,q} = (x_{p-1,q} + x_{p+1,q} +  x_{p,q+1} + x_{p,q-1})$, 
the individual terms of the Hamiltonian of the $(p, q)$th site can be written as  
\begin{eqnarray} 
\hat{h}_{p,q}^{{\rm TBH}} &=&  \sum_{\kappa} \bigg\{- J_\kappa\Big[ 
		 \big(\hat{b}_{p, q, \kappa}^{\dagger}\bar{\phi}_{p, q, \kappa} + {\rm H.c.}\big) -\phi_{p,q,\kappa}^{*}\bar{\phi}_{p, q, \kappa}\Big]  
       \nonumber\\
      &+& \frac{U_{\kappa}}{2}\hat{n}_{p, q,\kappa}
              \left (\hat{n}_{p, q,\kappa}-1\right)
              \bigg\}
                  + U_{12}\hat{n}_{p, q, 1} \hat{n}_{p, q, 2}\nonumber\\
      &-& \mu_\kappa\hat{n}_{p, q, \kappa} \nonumber\\    
\hat{h}_{p,q}^{{\rm DIT}} &=& \sum_{\kappa}\bigg\{T_{\kappa}
        \Big[\big(\hat{b}_{p, q, \kappa}^\dagger  
	      \hat{n}_{p, q, \kappa} \bar{\phi}_{p, q,\kappa}  
	    + \hat{b}_{p, q, \kappa}^{\dagger} 
	      \bar{\eta}_{p, q, \kappa} + {\rm H.c.}\big) 
          \nonumber\\
        &-& \eta_{p,q,\kappa}^{*} \bar{\phi}_{p, q,\kappa}
        -\phi_{p,q,\kappa}^{*}\bar{\eta}_{p,q,\kappa}\Big] 
        \nonumber\\  
	      &+& T_{\kappa, 3-\kappa} \Big[\big(\hat{b}_{p, q, \kappa}^{\dagger} \hat{n}_{p,q,3-\kappa}\bar{\phi}_{p, q, \kappa} 
	      + \hat{b}_{p, q, \kappa}^{\dagger} \bar{\eta}_{p, q, 3-\kappa, \kappa} \nonumber\\   
          &+& {\rm H.c.}\big) -\eta_{p,q,3-\kappa,\kappa}^{*} \bar{\phi}_{p, q, \kappa} - \phi_{p,q,\kappa}^{*}\bar{\eta}_{p, q, 3-\kappa, \kappa} \Big] \bigg\},\nonumber\\
\hat{h}_{p,q}^{{\rm NN}} &=& 
\sum_{\kappa} \bigg[ V_{\kappa} \big(\hat{n}_{p,q,\kappa}-\frac{\rho_{p,q,\kappa}}{2}\big)\bar{\rho}_{p, q, \kappa} \nonumber\\
&+&  V_{\kappa, 3-\kappa} \big(\hat{n}_{p, q, \kappa} - \frac{\rho_{p,q,\kappa}}{2}\big) \bar{\rho}_{p, q, 3-\kappa}\bigg],
\label{ham_ss_tbec_a}
\end{eqnarray}
where, the mean-fields are defined as $\phi_{p, q, \kappa} = \langle \hat{b}_{p, q, \kappa} \rangle $, $\rho_{p, q, \kappa} = \langle \hat{n}_{p, q, \kappa} \rangle$,  $\eta_{p, q, \kappa} = \langle\hat{n}_{p, q, \kappa} \hat{b}_{p, q, \kappa}\rangle$, and
$\eta_{p, q,  3-\kappa, \kappa} = \langle\hat{n}_{p, q, 3-\kappa}\hat{b}_{p, q,\kappa}\rangle$. 
The single-site Hamiltonian matrix is then constructed in the local (on-site) Fock basis and diagonalized to determine the ground state. Using this result, the ground state of the mean-field Hamiltonian in Eq.~(\ref{hamil_ss_tbec}) is obtained via the Gutzwiller ansatz \cite{jakub_05, yan_17, niederle_13, jaksch_02, damski_03_1, yan_17_1, yan_17_2, kaur_24}.

\subsection{Numerical methodology}
\label{sec_num}
In our numerical analysis, we obtain the ground state of the mean-field Hamiltonian self-consistently. For this purpose, we perform an iterative process where we construct the single site mean-field Hamiltonian matrix with respect to the local Fock basis, and then diagonalize it to obtain the ground state of a site. 
For bosonic particles, the site occupancy is unrestricted, but, in practice one can restrict the occupation number $m_\kappa\in [0,N_b -1]$ in Eq.~\ref{gw_2s}. The cutoff $N_b$ should be chosen sufficiently large than the average density $\sum_i\rho_{i,\kappa}/L^2$, which is in turn determined by the chemical potential $\mu_{\kappa}$. The single-site Hamiltonian matrix then becomes square matrix of dimension $N_b^2$.

In particular, we start with an initial guess values for the mean-field quantities in the Hamiltonian given in Eqs. (\ref{hamil_ss_tbec}) and (\ref{ham_ss_tbec_a}). After diagonalizing the matrix we update the mean-field quantities, of the site by new values computed using updated ground state as per Eqs. (\ref{gw_phi})-(\ref{gw_eta}). 
Next, we perform the similar procedure for the next site, where the updated mean-fields of the previous site have been used for computation. This procedure is carried out for all the sites in the system, which constitutes a sweep. In order to obtain a self-consistent solution the above sweep is carried out until an expected convergence with respect to the values of the mean-fields is obtained. It is to be noted that our implementation is site-dependent and does not reduce to an effective single-site theory. This allow us to simulate quantum phases having spatial inhomogeneity.  
We consider $L\times L$ lattice, where $L$ ranges from $10$ to $40$ , and the local occupation state of each component is restricted to $N_b = 10$. Thus, the single-site matrix dimension in our simulation is $N_b^2\times N_b^2  = 100 \times 100$. The above cut-off is justified as we we restrict parameter regime $\mu/U\leq 3$ throughout. The convergence criteria imposed to our simulation is $|\phi_{i,\kappa}^{l-1} - \phi_{i,\kappa}^l| \ll 10^{-6}$, where subscript $l$ denotes the $l$-th sweep. We employ periodic boundary conditions in both directions of the homogeneous 2D lattice to account the thermodynamic characteristics of the obtained quantum phases.
It is important to note that self-consistent mean-field methods can be prone to convergence toward local minima. When such solutions exist, the final state may depend on the initial guess. To mitigate this, we take rigorous steps to ensure that our solutions are not trapped in local minima. Specifically, we test a wide variety of initial conditions: uniform, structured, and random (uniform-distributed) coefficients in the Gutzwiller wavefunction, as well as different choices for the initial superfluid order parameter, including uniform and structured profiles. Furthermore, once a self-consistent solution is obtained, we perturb it slightly to verify its robustness. These thorough checks confirm that the quantum phases presented in the manuscript are not artifacts of convergence to spurious local minima.

\begin{figure}[ht!]
   \includegraphics[width=9.0cm]{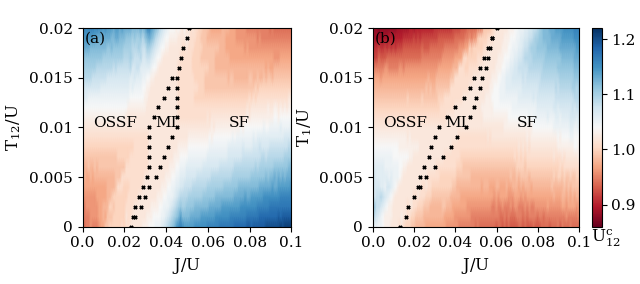}
   \caption{Dependence of $U^c_{12}$ on the DIT strengths, $T_1=T_2$ and $T_{12}$. We extend the parameter regime for our investigation by varying $J/U$ at a fixed chemical potential $\mu = 1.9 U$. With the increase $J/U$ parameter regions of OSSF, MI, and SF phases are explored.  (a)  fixed $T_{1} = T_{2} = 0.01 U$, but $T_{12}$ is varied. (b) fixed $T_{12} = T_{21} = 0.01 U$, but $T_{1}=T_{2}$ is varied. Data shows consistent behaviour as in the main text. Phase boundary for OSSF-MI-SF are plotted for $U_{12} = U $ and  to guide an eye to the reader.} 
  \label{ph_bou_t11_t12_j}
\end{figure}

\section{Influence of DIT processes on the MIT (varying $J$)}
\label{append_influ_dit}
Here, we extend our study on the dependence of the immiscibility condition by varying hopping 
strength $J$, and keeping chemical potential $\mu$ fixed.
As discussed earlier, critical value of $U^c_{12}$ for MIT depends 
on the inter and intra-component DIT processes for a fixed value of 
chemical potential and hopping strength in the phase diagram. 
To extend our results for larger parameter domain of the phase diagram, we vary the hopping strength for a fixed value of chemical potential, and we compute $U^c_{12}$. In the Fig.~\ref{ph_bou_t11_t12_j} (a), we show the heatmap of $U^c_{12}$ of MIT for chemical potential $\mu = 1.9U$, intra-component DIT, $T_{1} = 0.01U$.
As we increase hopping strength, we encounter three 
quantum phases OSSF, MI and SF. 
For the OSSF phase, we observe the enhancement in $U^c_{12}$ as we increase $T_{12}$. While in the SF phase, $U^c_{12}$ decreases with the increase in $T_{12}$. 
The black line in Fig.~\ref{ph_bou_t11_t12_j} marks the OSSF-MI and MI-SF phase boundaries.
In Fig.~\ref{ph_bou_t11_t12_j} (b), we show the heatmap for $U^c_{12}$ by fixing the inter-component DIT strength, $T_{12} = 0.01 U$. We now vary $T_{1} = T_{2}$ and  $J$ to perform similar analysis. Like earlier, we enter in the OSSF, MI and SF phases as 
we vary the hopping strengths. The dependence of $U^c_{12}$ shows opposite
trend compared to Fig.~\ref{ph_bou_t11_t12_j} (a). 
The critical value of $U^c_{12}$ decreases as we increase $T_{1}$ in the OSSF phase, 
while it increases in the SF phase. We emphasize that these observations are consistent with the results reported in the main text.

\begin{figure}[ht!]
   \includegraphics[width=9.0cm]{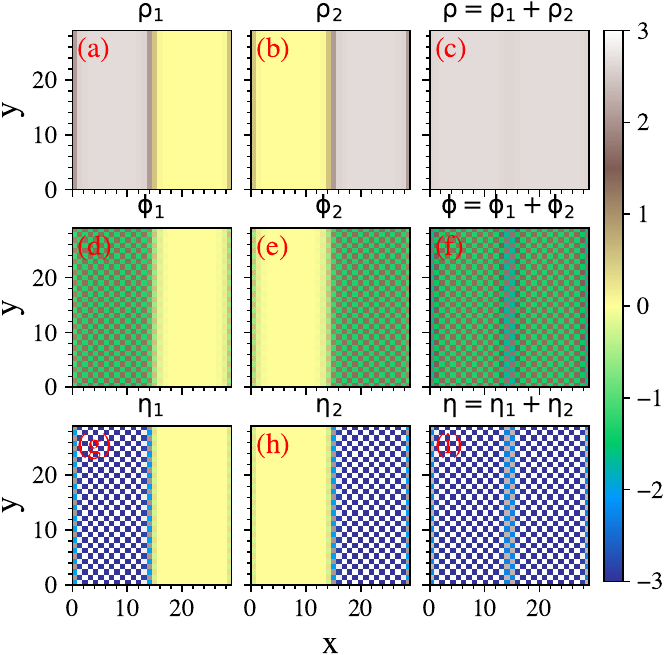}
   \caption{(a) - (c) Immiscible density distribution of the two-components in the OSSF phase. (a) and (b) illustrate the side-by-side homogeneous density profile of the components.
    The SF order parameter (d) - (e) and density transport 
    parameter (g) - (i) show sign staggering between adjacent sites and form a checkerboard pattern.
 	The considered parameters are $\mu = 1.9 U$, $J = 0.005 U$
    $U_{1} = U_{2} = U_{12} = 1$, $ T_{1} = T_{2} = 0.0117 U, T_{12} = 0.01$, $V_{1} = V_{2} = V_{12} =0$. Here the 
    critical value for the MIT is $T_{1} = T_{2} = 0.0116 U$. }
  \label{den_ossf}
\end{figure}
\begin{figure}[ht!]
  \includegraphics[width=9.0cm]{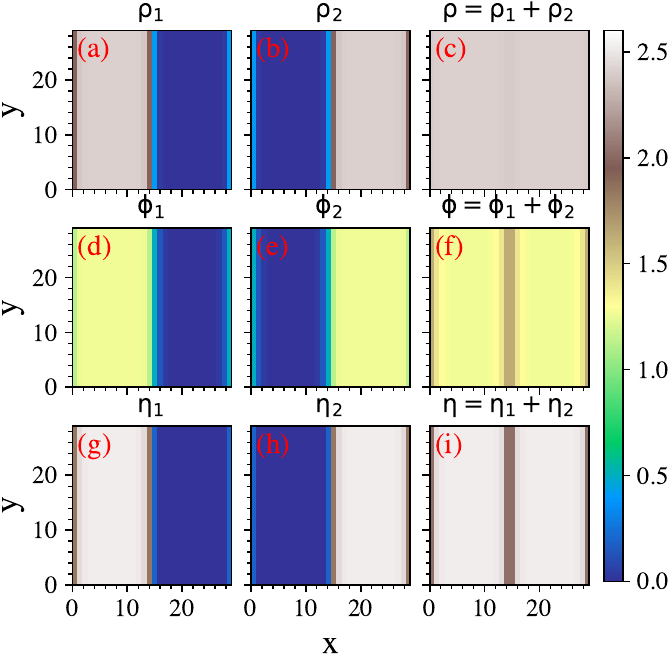}
  \caption{(a) - (c) Immiscible density distribution of the SF phase of the two-component BHM. Density distribution of the components are homogeneous and they form side-by-side profile.
    The SF order parameter (d) - (e) and density transport 
    parameter (g) - (i) are also homogeneous.
	The considered parameters are $\mu = 1.9 U$, $J = 0.1 U$,
    $U_{1} = U_{2} = U_{12} = 1$, $ T_{1} = T_{2} = 0.01 U, T_{12} = 0.0132$, $V_{1} = V_{2} = V_{12} =0$. Here the critical value for the MIT is $T_{12} = T_{21} = 0.0131 U$.}
  \label{den_sf}
\end{figure}

\section{Distributions of density and order parameters in the immiscible phase}

\begin{figure}[ht!]
  \includegraphics[width=9.0cm]{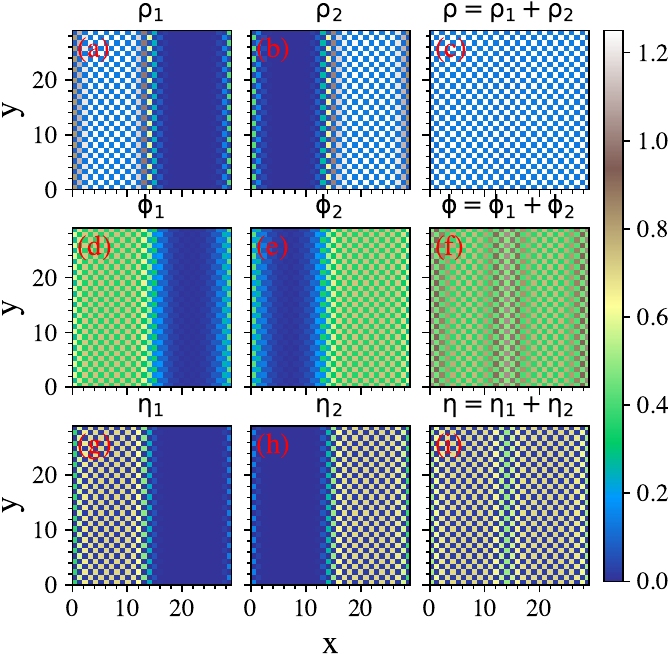}
  \caption{(a) - (c) Immiscible density distribution of SS 
    phase. In addition to forming side-by-side profile the density distribution of each component has checkerboard pattern.
    The SF order parameter (d) - (e) and density transport 
    parameter (g) - (i) also form a checkerboard pattern with real
    values between adjacent sites.
    The considered parameters are $\mu = 1.0 U$, $J = 0.08 U$,
    $U_{1} = U_{2} = U_{12} = 1$,
    $ T_{1} = T_{2} = 0.01 U, T_{12} = 0.0135$, $V_{1} = V_{2} = V_{12} =0.3 U$. Here the critical value for the MIT is 
    $T_{12} = T_{21} = 0.0134 U$.}
  \label{den_ss}
\end{figure}

Here, we present the density distributions of the immiscible phase in the OSSF, SF and SS
phase. To be specific, we show the side by side density distribution in these phases. 
For the OSSF phase of BHM, we consider $\mu = 1.9 U$, $J = 0.005 U$ and 
$T_1 = T_2= 0.0117U, T_{12} = T_{21} = 0.01 U$, and show the 
immiscible density distribution $\rho_\kappa$ in the Fig.~\ref{den_ossf} (a-c). 
In the Fig.~\ref{den_ossf} (d-f),
we show the corresponding SF order parameter $\phi_\kappa$ and in Fig.~\ref{den_ossf} (g-i) we show the corresponding density transport parameter $\eta_\kappa$. The SF order parameter and density transport parameter show sign staggering between adjacent site and forms a checkerboard pattern. Further, we observe the side by side density distributions in the SF and SS phase shown in Fig.~\ref{den_sf} and ~\ref{den_ss} respectively. 
For the SF phase, as mentioned earlier, the SF order parameter, density transport parameter and average density at each site is uniform. While in the SS phase, we observe that average density, SF order parameter and density transport parameter have checkerboard pattern. 
The numerical parameters for SF phase 
are $\mu = 1.9 U, J = 0.1 U$, and $T_1 = T_2= 0.01 U, T_{12} = T_{21} = 0.0132 U$. 
We set the NN interactions to be zero $V_{1} = V_{2} = V_{12} =0$ for both OSSF and SF phase.
While for SS phase we consider $V_1 = V_2 = V_{12} = 0.3U$, $\mu = 1.0U, J = 0.08U$ and 
$T_1 = T_2= 0.01 U, T_{12} = T_{21} = 0.0135 U$.

\section{Parameter estimation for the microscopic model}
\label{app_d}
Here, we outline how the experimentally feasible numerical parameters of the model considered in Eq. (\ref{hamil_terms}) can be calculated. In second quantization, the many-body Hamiltonian of a system consisting of a two-component bosonic mixture of same atomic species trapped in an optical lattice can be written as:
\begin{eqnarray}
&&\hspace{-0.35cm}\hat{H}_{\rm MB}= \sum_{\kappa}\int d\vec{r}_{1} \hat{\Psi}^{\dagger}_{\kappa}(\vec{r}_{1}) \left[-\frac{\hbar^2}{2m}\nabla^2 + V_{\text{ext}}(\vec{r}_{1})\right]
    \hat{\Psi}_{\kappa}(\vec{r}_{1}) \nonumber \\
   &&\hspace{-0.3cm}   + \frac{1}{2}\sum_{\kappa, \kappa^{\prime}} \int d\vec{r_1} d\vec{r_{2}} \hat{\Psi}^{\dagger}_{\kappa}(\vec{r}_{1}) \hat{\Psi}^{\dagger}_{\kappa^{\prime}}(\vec{r}_{2}) V_{\text{int}} (\vec{r}_{1},\vec{r}_{2}) \hat{\Psi}_{\kappa^{\prime}}(\vec{r}_{2}) \hat{\Psi}_{\kappa}(\vec{r}_{1})\nonumber
\end{eqnarray}
where $\kappa, \kappa^{\prime} \in {1,2}$ label the two components. The operators $\hat{\Psi}^{\dagger}_{\kappa}(\vec{r}_{1})$ and $\hat{\Psi}_{\kappa}(\vec{r}_{2})$ are bosonic creation and annihilation field operators for the $\kappa^{\text{th}}$ component, satisfying standard bosonic commutation relations. The external trapping potential $V_{\text{ext}}(\vec{r}_{1})$ includes the optical lattice potential, which in two dimensions takes the form 
\begin{equation}
V_{\rm OL} = V_{0} \left[\sin^2(\frac{\pi x}{a})+\sin^2(\frac{\pi y}{a})\right],
\end{equation}
corresponding to a $2$D square optical lattice with lattice constant $a$. 
The lattice depth $V_{0}$ is expressed in units of the recoil energy, defined as $E_{\rm R} = \hbar^2k^2/2m$, where the laser wavevector is $k=2\pi/\lambda =\pi/a$. The interaction energy between particles is represented by $V_{\text{int}} (\vec{r}_{1},\vec{r}_{2})$. This interaction arises from both short-range $V_{S}(r)$ and long-range $V_{L}(r)$ contributions:
\begin{equation}
    V_{\text{int}} (r) = V_{S}(r) + V_{L}(r) = g_{\kappa, \kappa^{'}}\delta(r) + \gamma_{\kappa, \kappa^{\prime}} h(r),
\end{equation}
where, $r = |\vec{r}_{1}-\vec{r}_{2}|$. Here, $g_{\kappa, \kappa^{\prime}}$ and $\gamma_{\kappa, \kappa^{\prime}}$ denote the strengths of the short-range (typically contact) and long-range interactions, respectively. The spatial form of the long-range potential $h(r)$ follows $1/r^3$ dependence for dipolar interactions.
The field operators can be expanded in the basis of Bloch waves, which can in turn be expressed using Wannier functions: 
\begin{equation}
    \hat{\Psi}_{\kappa}(\vec{r}) = \sum_{n,\vec{q}} \phi_{\kappa, n,\vec{q}}(\vec{r}) \hat{b}_{\kappa, n,\vec{q}}
    \end{equation}
with the Bloch functions given by:
\begin{equation}
    \phi_{\kappa, n,q}(\vec{r}) = \sum_{\vec{R}} \Omega_{\kappa,n,\vec{R}}(\vec{r}) e^{i\vec{q}.\vec{R}},
\end{equation}

where $\Omega_{\kappa,n,\vec{R}}(\vec{r})$ are the Wannier functions centered at lattice sites $\vec{R} = pa\vec{e}_x + qa\vec{e}_y$, $n$ denotes the band index, and $\vec{q}$ the quasi-momentum. As a result, we can also write:
\begin{equation}
    \hat{\Psi}_{\kappa}(\vec{r}) = \sum_{n,\vec{R}} \Omega_{\kappa,n,\vec{R}}(\vec{r}) \hat{b}_{\kappa, n,\vec{R}},
\end{equation}
where $\hat{b}_{\kappa, n,\vec{R}}$ are bosonic annihilation operators localized at site $\vec{R}$.

Working in the lowest-band approximation, one obtains Eq.~\ref{hamil_terms}, where the various parameters are expressed in terms of Wannier functions as follows. The tunneling amplitude is given by:
\begin{equation}
    J_{\kappa,i,j} =\int d\vec{r} \Omega^{*}_{\kappa, i}(\vec{r}) \left[ \frac{\hbar^2}{2m}\nabla^2 - V_{\text ext}(\vec{r})\right] \Omega_{\kappa, j}(\vec{r}),
\end{equation}
and the on-site interaction strength takes the form:
\begin{equation}
    U_{\kappa, \kappa^{\prime}} = g_{\kappa, \kappa^{\prime}} W^{S}_{iiii} + \gamma_{\kappa, \kappa^{\prime}} W^{L}_{iiii},
\end{equation}
where 
\begin{equation}
    W^{S}_{ijkl} = \int d\vec{r} \Omega^{*}_{\kappa, i}(\vec{r})  \Omega^{*}_{\kappa^{\prime}, j}(\vec{r})  \Omega_{\kappa, k}(\vec{r}) \Omega_{\kappa^{\prime}, l}(\vec{r}),
\end{equation}
and
\begin{align}
\begin{split}
   &W^{L}_{ijkl}\\
   &= \int d\vec{r}_{1} d\vec{r}_{2} \Omega^{*}_{\kappa, i}(\vec{r}_{1})  \Omega^{*}_{\kappa^{\prime}, j}(\vec{r}_{2}) h(|\vec{r}_{1}-\vec{r}_{2}|)  \Omega_{\kappa, k}(\vec{r}_{1}) \Omega_{\kappa^{\prime}, l}(\vec{r}_{2}).
\end{split}
\end{align}

To provide quantitative insight into the energy scales of the effective lattice Hamiltonian in Eq.~(\ref{hamil_terms}), we now estimate the relevant parameters—hopping amplitude $J$, $U$), density-density interaction $V$, and density-induced tunneling $T$—based on realistic experimental conditions.

For deep lattices \( (V_0 \gtrsim 5E_{\rm R})\), we can approximate the Wannier functions of the lowest band by Gaussians:
\begin{equation}
\Omega(\vec{r}) = \frac{1}{\sqrt{\pi} \ell} \exp\left(-\frac{r^2}{2\ell^2}\right),
\end{equation}
where the width $\ell$ scales as $\ell \sim a(V_0/E_{\rm R})^{-1/4}$.
The hopping amplitude between neighboring sites in the lowest band is well-approximated by~\cite{jaksch_98}:
\begin{equation}
J \approx \frac{4}{\sqrt{\pi}} E_{\rm R} \left(\frac{V_0}{E_{\rm R}}\right)^{3/4} \exp\left[-2\left(\frac{V_0}{E_{\rm R}}\right)^{1/2}\right].
\end{equation}
For example, with \( V_0 = 10E_{\rm R} \), this gives \( J \approx 0.02E_{\rm R} \).
The on-site interaction consists of a short-range contact contribution and a long-range dipolar part:

\begin{eqnarray}
U_{\kappa, \kappa^{\prime}} &&= g_{\kappa, \kappa^{\prime}} \int |\Omega(\vec{r})|^4 d^2r\nonumber\\
&&+ \gamma_{\kappa, \kappa^{\prime}} \int d\vec{r}_1 d\vec{r}_2 |\Omega(\vec{r}_1)|^2 |\Omega(\vec{r}_2)|^2 h(|\vec{r}_1 - \vec{r}_2|).\nonumber\\
\end{eqnarray}
Using the Gaussian approximation, the contact integral evaluates to:
\begin{equation}
\int |\Omega(\vec{r})|^4 d^2r = \frac{1}{2\pi \ell^2}.
\end{equation}
In quasi-two-dimensional setups with strong confinement along the \( z \)-direction (characterized by harmonic length \( \ell_z \)), the effective 2D coupling constant becomes \( g_{2D} \approx \sqrt{8\pi} \hbar^2 a_s / (m \ell_z) \), where \( a_s \) is the 3D \( s \)-wave scattering length. Thus, the contact interaction is:
\begin{equation}
U_{\rm contact} \approx \frac{\sqrt{8/\pi} \hbar^2 a_s}{m \ell_z \ell^2}.
\end{equation}
For typical parameters \( a_s = 100a_0 \), \( \ell \sim 0.3a \), and \( \ell_z \sim 0.1a \), we estimate \( U_{\rm contact} \approx 1.1 E_{\rm R} \).
The dipolar interaction strength is \( \gamma = C_{\rm dd}/(4\pi) \), with \( C_{\rm dd} = \mu_0 \mu^2 \) for magnetic dipoles (e.g., \( \mu = 10\mu_B \)). Assuming \( \ell \sim 150\,\text{nm} \), we obtain:
\begin{equation}
U_{\rm dip} \sim \frac{\gamma}{\ell^3} \sim 0.01 E_{\rm R},
\end{equation}
yielding a total on-site interaction \( U \sim 1.1 E_{\rm R} \).

Off-site density-density interactions arise purely from the long-range potential:
\begin{equation}
V_{\kappa,\kappa'} = \frac{\gamma_{\kappa,\kappa'}}{2} \left(W^{L}_{ijij} + W^{L}_{jiij}\right).
\end{equation}
These overlap integrals decay rapidly with distance, and their scaling is set by the dipolar strength and lattice spacing:
\begin{equation}
V \sim \frac{\gamma}{a^3}.
\end{equation}
Using \( a = 500\,\text{nm} \), we estimate \( V \sim 0.003 E_{\rm R} \).
Similarly, the DIT term is given by:
\begin{equation}
T_{\kappa,\kappa'} = \gamma_{\kappa,\kappa'} W^{L}_{ijjj},
\end{equation}
which depends on asymmetric overlap of Wannier functions centered on adjacent sites. It is typically smaller than \( V \), with an estimated value in the range:
\begin{equation}
T \sim 0.005 E_{\rm R}.
\end{equation}

In experiments with dipolar atoms, the off-site density-density interaction and density-induced tunneling (DIT) can be made dominant by changing the lattice spacing in an accordion optical lattice~\cite{LinSu_2023} and can also be tuned by adjusting the polarization direction of the dipoles~\cite{bandyopadhyay_19}. Furthermore, it has been shown that for atoms with predominantly on-site interactions in a tilted lattice, the DIT strength can become comparable to the hopping strength~\cite{Meinert_13, Jurgensen_14} by tuning the scattering length $a_s$ through Feshbach resonances~\cite{chin_10}.

\bibliography{ref}{}

\end{document}